\documentclass[journal]{IEEEtran}
\ifCLASSINFOpdf
\else
\fi
\usepackage{comment}
\usepackage{cite}
\usepackage{indentfirst}
\usepackage{graphicx}
\usepackage{changepage}
\usepackage{stfloats}
\usepackage{amsfonts,amssymb}
\usepackage{bm}
\usepackage{diagbox}
\usepackage{algorithm}
\usepackage{algorithmic}
\usepackage{amsmath}
\usepackage{amsmath,amssymb,amsfonts}
\usepackage{times}
\usepackage{url}
\usepackage{cite}
\usepackage{textcomp}
\usepackage{xcolor}
\usepackage{color}
\usepackage{setspace}
\usepackage{enumitem}
\usepackage{float}
\usepackage{diagbox}
\usepackage[T1]{fontenc}
\usepackage[utf8]{inputenc}
\usepackage{authblk}
\usepackage{threeparttable}
\usepackage{booktabs}
\usepackage{footnote}
\usepackage{graphicx}
\usepackage{pifont}
\usepackage{subfigure}
\usepackage{mathrsfs}
\usepackage{makecell}
\usepackage{array}
\usepackage{booktabs}
\usepackage{lipsum}
\usepackage{multirow}
\usepackage{adjustbox}
\usepackage{bbding}
\usepackage{tabularx}
\usepackage{colortbl}

\begin{document}
\abovedisplayshortskip=1pt
\belowdisplayshortskip=1pt
\abovedisplayskip=1pt
\belowdisplayskip=1pt
\textfloatsep=1pt
\floatsep=1pt
\intextsep=1pt
\setlength{\parskip}{0cm}
\setlength{\parindent}{1em}
%
\title{Network-Level ISAC Design: State-of-the-Art, Challenges, and Opportunities}
%
%
%

\author{Kawon Han,~\IEEEmembership{Member,~IEEE,}
        Kaitao Meng,~\IEEEmembership{Member,~IEEE,}
        Xiao-Yang Wang,~\IEEEmembership{Member,~IEEE,}
        and~Christos~Masouros,~\IEEEmembership{Fellow,~IEEE}

\thanks{Kawon Han, Kaitao Meng, and Christos Masouros are with the Department of Electronic and Electrical Engineering, University College London, London, UK (E-mail: {kawon.han, kaitao.meng, c.masouros}@ucl.ac.uk).}
\thanks{Xiao-Yang Wang is with the School of Information and Communication
Engineering, Beijing University of Posts and Telecommunications, Beijing
100876, China, also with the Key Laboratory of Universal Wireless
Communications, Ministry of Education, Beijing, China (E-mail: wangxy\_028@bupt.edu.cn).}}
\maketitle

\begin{abstract}
The ultimate goal of integrated sensing and communication (ISAC) deployment is to provide coordinated sensing and communication services at an unprecedented scale. This paper presents a comprehensive overview of network-level ISAC systems, an emerging paradigm that significantly extends the capabilities of link-level ISAC through distributed cooperation. We first examine recent advancements in network-level ISAC architectures, emphasizing various cooperation schemes and distributed system designs. The sensing and communication (S\&C) performance is analyzed with respect to interference management and cooperative S\&C, offering new insights into the design principles necessary for large-scale networked ISAC deployments. In addition, distributed signaling strategies across different levels of cooperation are reviewed, focusing on key performance metrics such as sensing accuracy and communication quality-of-service (QoS). Next, we explore the key challenges for practical deployment where the critical role of synchronization is also discussed, highlighting advanced over-the-air synchronization techniques specifically tailored for bi-static and distributed ISAC systems. Finally, open challenges and future research directions in network-level ISAC design are identified. The findings and discussions aim to serve as a foundational guideline for advancing scalable, high-performance, and resilient distributed ISAC systems in next-generation wireless networks.
\end{abstract}


%
\IEEEpeerreviewmaketitle

\section{Introduction}
\IEEEPARstart{I}{ntegrated} sensing and communication (ISAC) has emerged as a key technology for next-generation wireless systems, enabling the joint use of spectral and hardware resources for both radar sensing and data transmission \cite{Liu2022SurveyFundamental, meng2023sensing}. This convergence offers significant improvements in spectral and energy efficiency, while simultaneously reducing hardware complexity and deployment costs \cite{Cui2021Integrating}. Early research on ISAC primarily focused on link-level system designs \cite{Ouyang2022Performance, Hua20243DMultiTarget}, which, despite demonstrating the feasibility of integrating sensing and communication, revealed critical limitations: severe inter-cell interference, limited coverage in millimeter-wave bands, and restricted sensing resolution, accuracy, and detection probability \cite{Liu2022Integrated}.

To address these challenges, the paradigm of \emph{network-level ISAC}, also referred to as multi-cell cooperation, networked ISAC, cooperative ISAC, distributed ISAC, or cell-free ISAC in existing works, e.g., \cite{meng2024cooperative, guo2025integrated, li2025multi}, has been proposed. In this framework, multiple base stations (BSs) collaborate to enhance communication services through techniques such as resource allocation, coordinated beamforming (CBF), and joint transmission and reception via coordinated multi-point (CoMP) strategies. Simultaneously, distributed multi-input multi-output (MIMO) radar architectures introduce new opportunities within network-level ISAC by significantly improving target detection and localization performance.
Harnessing these opportunities necessitates a step change from today's communication-only coordination, which neither captures the opportunities for coordinated and multi-static sensing, nor addresses the conflicting communication and radar objectives.
Jointly designing sensing and communication functions across the network enables ISAC systems to achieve their ultimate goal: realizing coordinated sensing and communication at an unprecedented scale. Such cooperative strategies can expand coverage areas, mitigate inter-cell interference, and enrich spatial diversity, thereby supporting high-resolution sensing and robust communication coverage across diverse environments. 
 
\subsection{Opportunities of Network-Level ISAC}
\subsubsection{Hardware Perspectives}
As one of the primary goals of ISAC in 6G is to enable sensing capabilities within wireless mobile networks \cite{dong2022sensing}, it is essential to ensure reliable sensing performance to support a wide range of applications, including unmanned aerial vehicle (UAV) detection, vehicular monitoring, and robotic perception. However, achieving high radar sensing accuracy and resolution often necessitates increased hardware complexity, such as large antenna array apertures with wideband antennas RF chains, which may be impractical for large-scale or cost-sensitive deployments \cite{han2022high, schneider2021hybrid}. 

In particular, unlike conventional radar systems that typically employ frequency-modulated continuous wave (FMCW) waveforms, ISAC systems use digitally modulated waveforms designed primarily for data transmission. This shift imposes stringent hardware requirements, as analog-to-digital converters (ADCs) and digital-to-analog converters (DACs) must operate at very high sampling rates, often exceeding several gigasamples per second, to achieve centimeter-level range resolution \cite{lang2022ofdm, han2023sub, kang2024sub}. Furthermore, an ISAC receiver should be able to detect both a communication link, subject to single link attenuation, as well as a target echo which is subject to a much larger forward and reverse link attenuation. This has significant implications on the dynamic range of the receive low-noise amplifiers (LNAs) and ADCs. These constraints present significant design challenges for practical ISAC hardware implementation in mobile network environments \cite{schweizer2021fairy}. In this context, network-level ISAC presents a promising solution for achieving high-performance sensing without the need to increase the hardware specifications of individual ISAC nodes. By leveraging spatially distributed nodes, the system effectively forms a large aperture with sparsely sampled spatial points, enabling high-resolution and high-accuracy sensing through cooperative operation \cite{lehmann2006high}. 

On the other hand, the maximum transmission power of a typical RF chain is inherently limited by the linear output capabilities of RF power amplifiers (PAs), and the total system transmit power is further constrained by effective isotropic radiated power (EIRP) regulations. Consequently, the transmit power available at an individual ISAC node may be insufficient to support reliable dual-functional radar-communication operation, particularly under the influence of RF impairments such as transmitter and receiver nonlinearities \cite{kang2024sensing, michev2023enhancing}. This limitation naturally motivates the adoption of network-level cooperative ISAC systems, where distributed nodes jointly transmit and receive to provide a transmit power combining gain and mitigate individual hardware constraints \cite{ji2023networking}. Therefore, the network-level distributed ISAC system opens up new opportunities to relax stringent hardware requirements at the node level, including reductions in antenna array aperture, ADC/DAC sampling rates, RF front-end bandwidth, and per-node transmission power.

\subsubsection{Application Perspectives}
Building upon the concept of network‐level ISAC, a wide range of opportunities arise for deploying ISAC systems across diverse scenarios \cite{gonzalez2025six}. Network-level ISAC principles extend to space-air-ground integrated (SAGI) networks, which leverage satellites, aerial nodes, and terrestrial stations to collect and fuse multi‐modal sensory data, including radar, LiDAR, cameras, and GPS, facilitating dynamic resource allocation and comprehensive environmental awareness, while enabling cooperative gains between non-terrestrial and terrestrial infrastructures \cite{sallouha2024ground,jiang2025network}. Moreover, the synergy between networked sensing and communication enhances coverage resilience and service continuity across varied deployment environments \cite{manzoni2024wavefield}. In vehicular ISAC, collaborative sensing among vehicles augments situational awareness for traffic coordination and platooning, and dedicated MAC protocols alongside real-time AI-driven updates optimize latency and resource management. For example, the distributed intelligent sensing and communication architecture introduces edge-distributed processing and semantic-level data fusion to support seamless target handovers and scalable low-latency S\&C services in densely deployed networks \cite{strinati2025toward,ge2024target}. As AI becomes a central 6G pillar, coordinated sensing with real-time data collection across heterogeneous nodes underpins self-adaptive network intelligence, enabling dynamic protocol adjustment and predictive configuration.

\subsection{Organization of the Paper}
While network-level ISAC offers promising opportunities for scalable ISAC deployment, implementing multi-cell cooperation introduces several challenges, including increased backhaul and signaling overhead, complex joint resource allocation across antennas, bandwidth, and processing units, as well as stringent synchronization requirements among spatially distributed nodes. To fully exploit the potential of distributed ISAC, it is essential to develop comprehensive system architectures alongside efficient signaling and robust synchronization strategies that adapt to varying cooperation levels and practical system constraints. 

In light of these challenges, this paper provides a comprehensive overview of recent advances on network-level ISAC design, spanning stochastic geometry (SG)-based performance analysis, distributed ISAC signaling design, and synchronization techniques. Section~\ref{sec::Architecture} reviews cooperation schemes and distributed architectures for networked ISAC, including fundamental transmitter and receiver system models. Section~\ref{sec::Performance} presents the performance analysis and scaling laws of network-level ISAC, offering new insights into efficient system design. Subsequently, Section~\ref{sec::Signaling} discusses distributed ISAC signaling strategies based on various signaling models. The critical role of synchronization, particularly over-the-air synchronization techniques, is explored in Section~\ref{sec::Sync}. Finally, remaining challenges and future research directions for network-level ISAC design are outlined in Section~\ref{sec::Challenge}.

$Notations$: Boldface variables with lower- and upper-case symbols represent vectors and matrices, respectively. $\textbf{A} \in \mathbb{C}^{N \times M}$ and $\textbf{B} \in \mathbb{R}^{N \times M}$ denotes a complex-valued ${N \times M}$ matrix $\textbf{A}$ and a real-valued ${N \times M}$ matrix $\textbf{B}$, respectively. Also, $\mathbf{0}_{N \times M}$ and $\mathbf{I}_N$ denote a $N \times M$ zero-matrix and a $N \times N$ identity matrix, respectively. $({\cdot})^{T}$, $({\cdot})^{H}$, and $({\cdot})^{*}$ represent the transpose, Hermitian transpose, and conjugate operators, respectively. $\mathbb{E}{[\cdot]}$ is the statistical expectation operator.

\begin{figure}[t!]
	\centering
	\includegraphics[width=8.4cm]{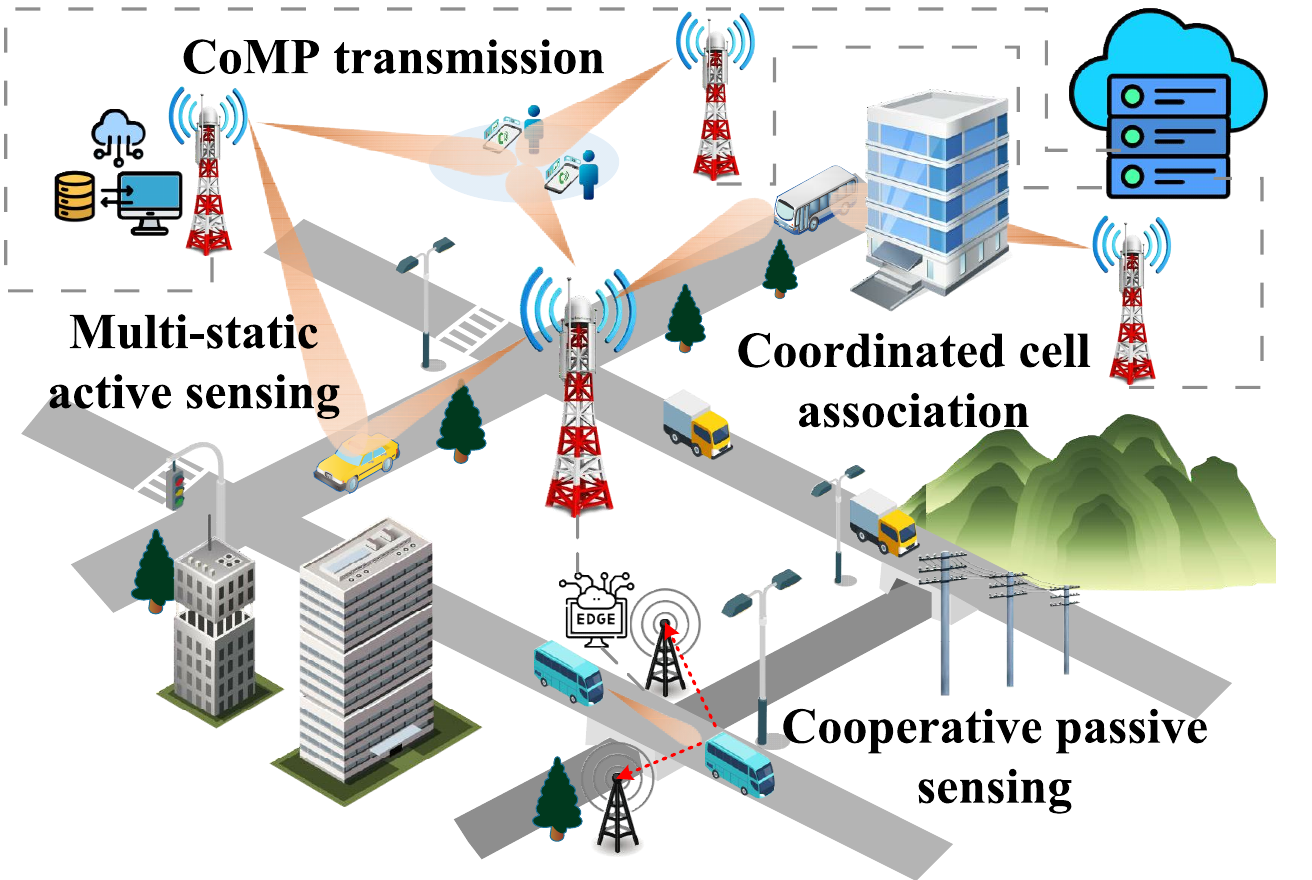}
	\caption{Scenarios of network-level distributed ISAC.}
	\label{figure1}
\end{figure}

\begin{table*}[]
  \centering
  \footnotesize
  \vspace{-3mm}
  \caption{Comparison of different ISAC cooperation networks and requirements}
  \label{Table2_updated6}
  \begin{tabular}{|l|l|l|l|l|l|l|}
    \hline
    \multicolumn{2}{|l|}{\diagbox[width=4.5cm]{\bf{Cooperation Level}}{\bf{Requirement}}}
      & {\bf{Information Sharing}}
      & {\bf{Network Connectivity}}
      & {\makecell[c]{\bf{Synchronization}}}
      & {\makecell[c]{\bf{Signaling} \\ \bf{Overhead}}}
      & {\makecell[c]{\bf{Cooperation} \\ \bf{Gain}}} \\
    \hline

    \multicolumn{2}{|l|}{Non-cooperative (Baseline)}
      & None
      & Singleton links
      & None
      & Low
      & None
    \\ \hline

    \multicolumn{2}{|l|}{Interference Management}
      & User/target CSI
      & Pairwise BS links
      & None
      & Medium
      & High
    \\ \hline

    \multirow{2}{*}{Decentralized Cooperation}
      & Non-coherent
      & Data/echo signals
      & Clustered neighbors
      & time, frequency
      & Very high
      & High
    \\ \cline{2-7}
      & Coherent
      & CSI + Data/echo signals
      & Clustered neighbors
      & time, frequency, phase
      & Very high
      & Very high
    \\ \hline

    \multirow{2}{*}{Centralized Cooperation}
      & Non-coherent
      & Global CSI + All data
      & Full connectivity
      & time, frequency
      & Extreme
      & Extreme
    \\ \cline{2-7}
      & Coherent
      & Global CSI + All data
      & Full connectivity
      & time, frequency, phase
      & Extreme
      & Extreme
    \\ \hline

  \end{tabular}
\end{table*}

\section{Architecture and System Models of Network-Level Distributed ISAC} \label{sec::Architecture}
\subsection{Cooperation Schemes in Network-Level ISAC}
In ISAC networks, cooperation can be structured along a spectrum of architectural layers, each trading off complexity and overhead for improved sensing and communication performance.  At the simplest extreme, non‐cooperative operation (no CSI or data exchange, single‐link connectivity) imposes minimal signaling cost but yields no cooperation gain \cite{10320397}.  Introducing interference management, where only user/target CSI is shared over pairwise BS links, allows basic beamforming nulls to suppress dominant interferers without synchronization, delivering a modest uplift in area spectral efficiency and sensing accuracy \cite{10735119}.  

Stepping up to decentralized cooperation, clusters of neighboring BSs pool raw echo and/or CSI to perform either non‐coherent (symbol‐level aligned) or coherent (phase‐aligned) joint beamforming.  Here, the enriched information sharing and tighter synchronization unleash very high signaling overhead and computational burden, but in return achieve substantially higher S\&C gains via distributed MIMO diversity and nulling \cite{Meng2024CooperativeTWC}. Centralized cooperation funnels global CSI and all data into a fusion center over a full‐mesh backhaul: the non‐coherent fusion mode relaxes timing constraints to symbol‐level, while coherent fusion demands network‐wide phase synchronization.  Although this configuration incurs extreme overhead, it maximally exploits spatial degrees of freedom, enabling near–theoretical upper‐bound performance in both communication throughput and positioning accuracy (see Table~\ref{Table2_updated6}).

\begin{figure}[t]
	\centering
	\includegraphics[width=8.9cm]{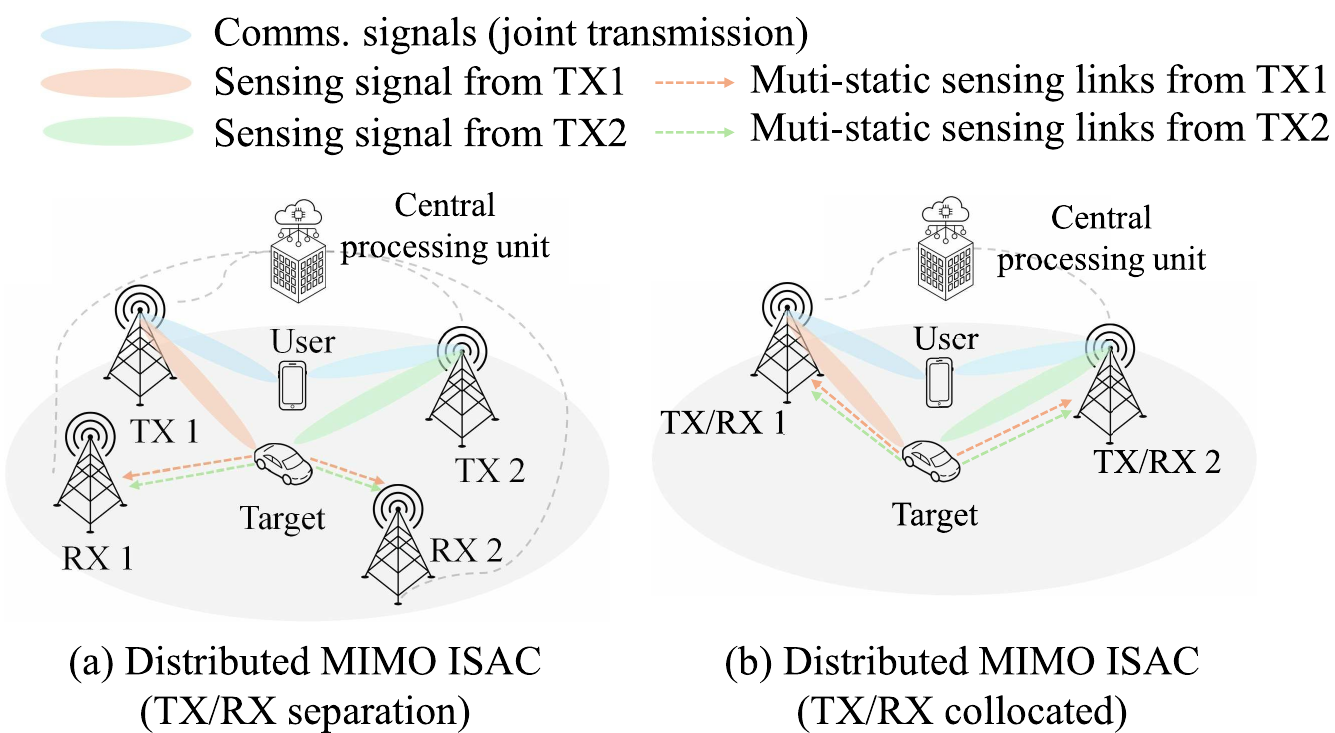}
	\caption{Distributed MIMO ISAC: (a) TX/RX separated architecture with passive multi-static sensing, and (b) TX/RX collocated architecture with hybrid of mono-static and bi-static sensing.}
	\label{Fig_KW1}
\end{figure}
\subsection{Distributed ISAC Architectures}
Distributed ISAC systems exploit spatial diversity by employing widely separated transmitters, receivers, or both. To this end, various configurations, such as multi-input single-output (MISO), single-input multi-output (SIMO), and multi-input multi-output (MIMO), can be adopted, each presenting distinct trade-offs in terms of system complexity, cooperative gain, and transceiver design. The studies in \cite{zhang2024target, wei2024integrated, chowdary2024hybrid} investigate a cooperative ISAC system consisting of multiple transmitters and a single receiver. While this configuration enables distributed transmission gain, it inherently lacks receiver diversity from the radar sensing perspective. Furthermore, coordination among distributed transmitters is crucial for interference nulling or joint transmission to support wireless communication. Notably, this setup simplifies receiver-side processing, as it eliminates the need for signal fusion due to the presence of only a single receiver.

On the other hand, a distributed SIMO system leverages receiver diversity, wherein widely separated multi-static receivers collaboratively process the received signals to enhance target detection and parameter estimation performance \cite{lou2024beamforming, jiang2024cooperation}. Receiver-side processing for radar sensing must account for the degree of data fusion across the distributed receivers to achieve cooperative sensing gain. Decentralized processing, which relies on fusing local estimates, reduces the backhaul burden by minimizing the need for sharing raw data. Nevertheless, it cannot achieve coherent processing gain for tasks such as target localization and velocity estimation, resulting in suboptimal performance compared to centralized approaches. In contrast, centralized processing enables optimal receiver performance through joint processing of raw data, but imposes significant demands on communication infrastructure for data aggregation. Therefore, the design of distributed ISAC systems with multiple receivers must carefully balance processing complexity and the achievable cooperative gain.    

The distributed MIMO configuration employs both multiple transmitters and multiple receivers to fully exploit cooperative gains \cite{yang2024coordinated, ji2023networking, wang2024collaborative, han2025signaling}. This setup necessitates coordinated transmission as well as data fusion across multiple receivers. Notably, distributed MIMO ISAC can be implemented in two distinct configurations, as illustrated in Fig.~\ref{Fig_KW1}: a half-duplex (HD) architecture in which transmitters and receivers are fully separated, and a full-duplex (FD) configuration where collocated transceivers are spatially distributed across the network. While both configurations require similar coordination mechanisms for communication, their sensing architectures differ. The fully separated setup forms a multistatic sensing system composed of multiple bi-static links, whereas the collocated transceiver setup combines mono-static and bi-static sensing links. These structural differences lead to distinct sensing performance characteristics, particularly in terms of synchronization requirements and sensitivity to phase noise, as mono-static and bi-static sensing links are differently affected by these impairments. However, the FD operation inherently introduces transmitter self-interference (SI) to the radar receivers, necessitating effective SI suppression techniques to preserve sensing performance \cite{barneto2019full}. The corresponding signal model and mitigating solutions will be detailed in Section~\ref{SensingModel}.

\subsection{Transmitter System Model}
In this section, we present a transmit signal model for ISAC in distributed and networked systems. Due to the larger power budget available at the BS necessary for sensing, our focus is on downlink transmission. Unlike collocated single-node ISAC signaling, distributed ISAC signaling depends heavily on the coordination strategy employed among multiple nodes. Assuming that each node is equipped with $M_t$ transmit antennas, we describe two signaling strategies: CBF and joint transmission CoMP.
As a general ISAC signal model for cooperative ISAC systems, the transmit signal at node $n$, $\forall n = 1, 2, \ldots, N$, is expressed as
\begin{align}
    \mathbf{X}_n = \sqrt{\rho^{c}} \mathbf{W}_n^{c} \mathbf{S}_n^{c} + \sqrt{\rho^{s}} \mathbf{W}_n^{s} \mathbf{S}_n^{s},
\end{align}
where $\rho^{c}$ and $\rho^{s}$ denote the transmit powers allocated to communication and sensing signals, respectively. The matrices $\mathbf{W}_n^{c} \in \mathbb{C}^{M_t \times K}$ and $\mathbf{W}_n^{s} \in \mathbb{C}^{M_t \times M_t}$ represent the corresponding precoding matrices for $K$ users and $U$ targets. $\mathbf{S}_n^{c} \in \mathbb{C}^{K \times L}$ and $\mathbf{S}_n^{s} \in \mathbb{C}^{M_t \times L}$ denote the modulated communication signals and sensing signals, respectively, with the block length of $L$. Note that by setting $\rho_c$ or $\rho_s$ to zero, the above model results into a joint ISAC transmission with a single signal component serving both sensing and communication \cite{liu2018toward}.

In CBF, which primarily mitigates inter-node interference for communication, each base station serves a distinct set of users. As a result, the communication signals from different nodes are uncorrelated, satisfying
\begin{align}
    \mathbb{E}[\mathbf{S}_n^{c} (\mathbf{S}_m^{c})^{H}] = \mathbf{0}_{M_t \times M_t}, \quad \forall n \neq m.
\end{align}
On the other hand, joint transmission CoMP involves multiple cooperating BSs transmitting the same data symbols to a typical user. In this case, the transmitted communication signals are identical across nodes, satisfying
\begin{align}
    \mathbb{E}[\mathbf{S}_n^{c} (\mathbf{S}_m^{c})^{H}] = \mathbf{I}_{M_t \times M_t}, \quad \forall n \neq m,
\end{align}
which implies $\mathbf{S}_1^{c} = \mathbf{S}_2^{c} = \cdots = \mathbf{S}_N^{c}$. The sensing signal $\mathbf{S}_n^{s}$ and its associated beamforming matrix $\mathbf{W}_n^{s}$ are designed to ensure the desired distributed sensing performance, considering target detection, parameter estimation accuracy, and coordination across nodes.

\subsection{Receiver System Model} \label{SensingModel}
\subsubsection{Communication Receiver Model}
Based on the transmitter models for CBF and joint transmission CoMP, the received signal at a typical user $k$, equipped with a single antenna, can be generally expressed as
\begin{equation}
    \mathbf{y}_{c,k}= \sum\nolimits_{n=1}^{N}\left\|\mathbf{d}_{k,n}\right\|^{-\frac{\alpha}{2}} \mathbf{h}_{k, n}^H \mathbf{X}_n + \mathbf{z}_{c,k}, \label{comm_sig1}
\end{equation}
where $\left\|\mathbf{d}_{k,n}\right\|$ denotes the distance between node $n$ and user $k$, $\alpha \geq 2$ is the pathloss exponent, and $\mathbf{h}_{k,n}^H \sim \mathcal{CN}\left(0, \mathbf{I}_{M_t}\right)$ represents the channel vector from node $n$ to user $k$. The term $\mathbf{z}_{c,k}$ denotes the additive noise at the receiver. The desired communication signal depends on the adopted transmitter coordination scheme. 

In the case of CBF, where each user is served by a designated node and transmissions from other nodes are treated as interference, the received signal in \eqref{comm_sig1}, assuming that user $k$ is served by node 1 and each node serves $K$ different users, can be reformulated as
\begin{equation}
	\begin{aligned}
		\mathbf{y}_{c,k}= \underbrace{ \tilde{\mathbf{h}}_{k, 1}^H \mathbf{w}^{c}_{k,1} \mathbf{s}^{c}_{k,1}}_{\text{intended signal}} & + \underbrace{\tilde{\mathbf{h}}_{k, 1}^H (\mathbf{X}_1 - \mathbf{w}^{c}_{k,1} \mathbf{s}^{c}_{k,1})}_{\text{multi-user + radar interference}}\\
        &+ \underbrace{\sum\nolimits_{n=2}^{N}\tilde{\mathbf{h}}_{k, n}^H \mathbf{X}_n}_{\text{inter-node interference}} + \mathbf{z}_{c,k},
	\end{aligned} \label{comm_coordinated}
\end{equation}
where we define $\tilde{\mathbf{h}}_{k, n}$ as the channel vector including the pathloss. Here, the inter-node interference acts as additive noise to the desired signal, emphasizing the need for coordination among ISAC nodes to manage interference effectively.

In contrast, the joint transmission CoMP strategy involves all cooperative base stations transmitting the same data stream to the user. Assuming that all $N$ nodes cooperatively serve $K$ users, the received signal at the typical user $k$ is then given by
\begin{equation}
	\begin{aligned}   \mathbf{y}_{c,k}=&\underbrace{\sum\nolimits_{n=1}^{N}\tilde{\mathbf{h}}_{k, n}^H \mathbf{w}^c_{k,n} \mathbf{s}^c_{k}}_{\text{collaborative intended signal}} \\
        &+ \underbrace{\sum\nolimits_{n=1}^{N}\tilde{\mathbf{h}}_{k, n}^H (\mathbf{X}_n -\mathbf{w}^c_{k,n} \mathbf{s}^c_{k})}_{\text{multi-user + radar interference}} + \mathbf{z}_{c,k}.
	\end{aligned} \label{comm_CoMP}
\end{equation}
In this case, the inter-node interference observed in CBF is transformed into a collaborative signal component that constructively enhances the received signal power and improves communication performance. However, the gain achievable through joint transmission is fundamentally limited by the level of phase synchronization among distributed nodes. Additionally, this approach requires significantly higher backhaul capacity to share the common data stream across all cooperating base stations. For further details on signal modeling including inter-cluster interference, the reader is referred to \cite{10735119, Meng2024CooperativeTWC}.

\subsubsection{Sensing Receiver Model} 
Assuming that each ISAC BS is equipped with $M_r$ receive antennas, the signal component relating to the target echo received at node $n$ from all transmitting nodes and reflected by target $u$ can be expressed as
\begin{align}
    \mathbf{Y}_{s,n}^{u} = \sum_{m=1}^{N} \left( \beta_{n,m}^{u} \mathbf{a}_{r,n}(\theta_{n}^{u}) \left( \mathbf{a}_{t,m}^T(\theta_{m}^{u}) \mathbf{X}_m \mathbf{J}_{\tau_{n,m}^{u}} \right) \right), \label{Eqn::8}
\end{align}
where $\mathbf{a}_{r,n}(\theta_{n}^{u}) \in \mathbb{C}^{M_r \times 1}$ and $\mathbf{a}_{t,m}(\theta_{m}^{u}) \in \mathbb{C}^{M_t \times 1}$ denote the receive and transmit steering vectors at nodes $n$ and $m$, respectively. The coefficient $\beta_{n,m}^{u}$ represents the complex amplitude incorporating radar cross-section (RCS) and propagation path loss. The matrix $\mathbf{J}_{\tau_{n,m}^{u}}$ denotes the temporal shift corresponding to the round-trip delay $\tau_{n,m}^{u}$ of the echo reflected from target $u$, following the formulation in \cite{liu2020range}.

The total signal received at node $n$, including SI, inter-node interference (INI), and reflections from all $U$ targets, is expressed as
\begin{align}
    \mathbf{Y}_{s,n} = \underbrace{\sum_{u=1}^{U} \mathbf{Y}_{s,n}^{u}}_{\text{target reflections}} + \underbrace{\sum_{m=1}^{N} \mathbf{Y}_{\text{INI},n,m}}_{\text{INI}} + \underbrace{\mathbf{Y}_{\text{SI},n}}_{\text{SI}} + \mathbf{Z}_{s,n}, \label{Eqn::9}
\end{align}
where $\mathbf{Y}_{\text{INI},n,m} = \mathbf{H}_{\text{INI},n,m}^H \mathbf{X}_m$ represents the direct INI from node $m$ to node $n$, with $\mathbf{H}_{\text{INI},n,m}$ denoting the INI channel matrix. The term $\mathbf{Y}_{\text{SI},n} = \mathbf{H}_{\text{SI},n}^H \mathbf{X}_n$ denotes the SI at node $n$, where $\mathbf{H}_{\text{SI},n} \in \mathbb{C}^{M_t \times M_r}$ is the SI channel matrix. Finally, $\mathbf{Z}_{s,n}$ denotes the additive noise at the sensing receiver of node $n$. The direct INI can be mitigated through interference nulling techniques \cite{10735119} or canceled using prior knowledge of the INI channels.

Importantly, the fully separated architecture of distributed ISAC, illustrated in Fig.~\ref{Fig_KW1}(a), operates in a HD mode and is inherently free from SI. In contrast, the FD distributed MIMO configuration in Fig.~\ref{Fig_KW1}(b) is subject to SI, which can significantly degrade sensing performance if not properly mitigated in both the RF and digital domains \cite{barneto2019full}. Digital SI cancellation techniques at the radar receiver, such as CLEAN-like algorithms and adaptive filtering, have been proposed to suppress SI in the baseband domain \cite{zeng2018joint, tang2024interference}. However, RF-domain cancellation remains critical to prevent the saturation of the LNA, necessitating more complex hardware implementations. These include RF leakage cancellation circuits employing multi-tap RF delay lines and vector modulators \cite{biedka2019full, huusari2015wideband}.

Based on the level of signal fusion, receiver processing in distributed ISAC systems can be categorized into two main schemes: centralized and decentralized processing. In centralized processing, a central processing unit (CPU) collects raw sensing data from all $N$ receiving nodes to perform joint target detection and parameter estimation. The collective received signal is represented as
\begin{align}
    \mathbf{Y}_{s} = \left[\mathbf{Y}_{s,1}^T, \mathbf{Y}_{s,2}^T, \ldots, \mathbf{Y}_{s,N}^T\right]^T,
\end{align}
where the goal is to estimate the target parameters using the aggregated signal observations from all nodes. This approach enables coherent processing when the phase-level synchronization is achieved and typically yields optimal performance in terms of sensing accuracy. In contrast, decentralized processing performs local target detection and estimation at each node using $\mathbf{Y}_{s,n}$, $\forall n$. The resulting local estimates are then fused at the CPU to produce a refined output. While decentralized processing significantly reduces backhaul and computational overhead compared to its centralized counterpart, it generally yields suboptimal performance due to the lack of coherent gain and is more susceptible to hardware impairments, such as synchronization errors and phase noise.

\section{Performance Analysis for Interference Management and Cooperative ISAC} \label{sec::Performance}

Recent advances leveraging stochastic geometry (SG) and information-theoretic analyses have begun to characterize the fundamental trade-offs in network-level ISAC \cite{10320397}, particularly among area spectral efficiency, coverage probability, and localization accuracy, under constraints on spatial degrees of freedom. To effectively address these trade-offs, unified performance metrics, such as joint coverage probability, ergodic rate, and potential spectral efficiency, need to be carefully explored for guiding the design of adaptive clustering, power control, and scheduling or joint resource allocation strategies in network-level ISAC systems.

As summarized in Table \ref{TablePerformanceMetric}, various metrics can be used to assess ISAC network performance. Service coverage probability \(P_x\)\cite{10562219,10569084,10633859,9420372,10568512,10433485,10320397} and localizability probability \(P_L\)\cite{10556618} quantify link availability under SINR  constraints, while ergodic rate \(R_x\)\cite{10490156} and potential spectral efficiency \(R_x^{\mathrm{PSE}}\)\cite{10490156} reflect average and fixed-threshold throughput performance in fading and threshold-based regimes. Area spectral efficiency \(T_x^{\mathrm{ASE}}\)\cite{10735119} and energy efficiency (EE)\cite{10490156} evaluate spatial and power resource utilization by measuring bits per unit area and per unit energy, respectively. Finally, ISAC coverage\cite{10320397,10615428} and ISAC ergodic rate\cite{10320397} provide weighted joint metrics that seamlessly integrate communication and sensing performance under varying network densities.

\begin{table*}[htbp] 
\centering
\footnotesize
\caption{Categorized performance metrics for network-level ISAC design}
\label{TablePerformanceMetric}
\renewcommand{\arraystretch}{1.3}
\begin{tabular}{|p{3.2cm}|p{4cm}|p{4.5cm}|p{4.5cm}|}
\hline
\textbf{Category} & \textbf{Metric} & \textbf{Mathematical Expression} & \textbf{Description / Reference} \\
\hline

\multirow{4}{*}{\textbf{Coverage Probability}} 
& Sensing or communication coverage & 
$\displaystyle P_x = \Pr[\mathrm{SINR}_x > \gamma_x],\quad x\in\{\mathrm{com},\mathrm{sen}\}$ & 
Communication or sensing SINR exceeding threshold $\gamma_x$ \cite{10562219,10569084,10633859,9420372,10568512,10433485,10320397}. \\
\cline{2-4}
& ISAC coverage & 
\small{$\frac{\lambda_\mathrm{UE} \mathbb{P}^{0}_{\Phi_\mathrm{UE}}[\mathrm{SINR_{com}}\geq \gamma_{\mathrm{com}}]}{\lambda_\mathrm{UE}+\lambda_\mathrm{s}} + \frac{\lambda_\mathrm{s} \mathbb{P}^{0}_{\Phi_\mathrm{s}}[\mathrm{SINR_{sen}}\geq \gamma_{\mathrm{sen}}]}{\lambda_\mathrm{UE}+\lambda_\mathrm{s}}$} & 
Weighted joint coverage based on user/target density \cite{10320397,10615428}. \\
\cline{2-4}
& Joint localization and communication coverage & 
$\displaystyle P_{p\&c}(\epsilon,\epsilon_c)
= \Pr[{\rm{CRLB}}\le\epsilon,\;\mathrm{SINR}_{\mathrm{com}}\ge\epsilon_c]$ & 
Joint probability that localization error and communication SINR meet required thresholds. \\
\cline{2-4}
& $\mathcal{L}$-Localizability Probability & 
\small{$P_L(\mathcal{L}|\gamma) = \mathbb{P}\left[\frac{P_t h_\mathcal{L} r_\mathcal{L}^{-\beta}}{\sum_{i=\mathcal{L}+1}^{\infty}P_t h_i r_i^{-\beta}+N_0} \geq \gamma\right]$} & 
Probability that top $\mathcal{L}$ strongest BSs meet localization SINR requirement \cite{10556618}. \\
\hline

\multirow{4}{*}{\textbf{Spectral Efficiency}} 
& Ergodic rate & 
$\displaystyle R_x = \mathbb{E}\bigl[\log(1 + \mathrm{SINR}_x)\bigr]$ & 
Average data rate (com) or radar info rate (sen) \cite{10490156}. \\
\cline{2-4}
& Potential spectral efficiency & 
$\displaystyle R_x^{\mathrm{PSE}} = \lambda_{bs}\,\log_2(1+\gamma_x)\,P_x$ & 
SE estimate based on SINR threshold and coverage probability \cite{10490156}. \\
& ISAC ergodic rate & 
$\displaystyle R_{\mathrm{ISAC}}
= \frac{\lambda_{UE}\,R_{\mathrm{com}} + \lambda_{s}\,R_{\mathrm{sen}}}
       {\lambda_{UE}+\lambda_{s}}$ & 
Density-weighted joint ergodic rate for comm and sensing \cite{10320397}. \\
\cline{2-4}
& Area spectral efficiency & 
$\displaystyle T_x^{\mathrm{ASE}} = \lambda_{bs}\,N_x\,R_x,\quad N_x=\begin{cases}
K,&x=\mathrm{com}\\
J,&x=\mathrm{sen}
\end{cases}$ & 
Spectral efficiency per unit area; $K$ users or $J$ targets per cell \cite{10735119}.  \\
\hline

\multirow{1}{*}{\textbf{Estimation accuracy}} 
& CRLB, \text{MSE}, etc. & 
${\rm{E}}\left[\operatorname{tr}\left( {\mathbf{F}}^{-1}\right)\right]$, 
$\mathbb{E}  \left\{  (\hat{\theta} - \theta )^{2}  \right\}$ & 
Average lower bound of parameter estimation error \cite{Meng2024CooperativeTWC, xiong2017fda}. \\
\hline

\multirow{1}{*}{\textbf{Energy Efficiency}} 
& Information rate over power consumption & 
$\displaystyle \mathrm{EE} \!=\! \frac{R_{\mathrm{com}} \!+\! R_{\mathrm{sen}}}{P_\textrm{hardware} \!+\! P_{\textrm{transmit}} + P_{\textrm{mechanic}} }$ & 
Sum rate divided by system power consumption \cite{10490156, virgili2022cost}. \\
\hline
\end{tabular}
\end{table*}

\begin{figure}[t]
	\centering
	\includegraphics[width=8.4cm]{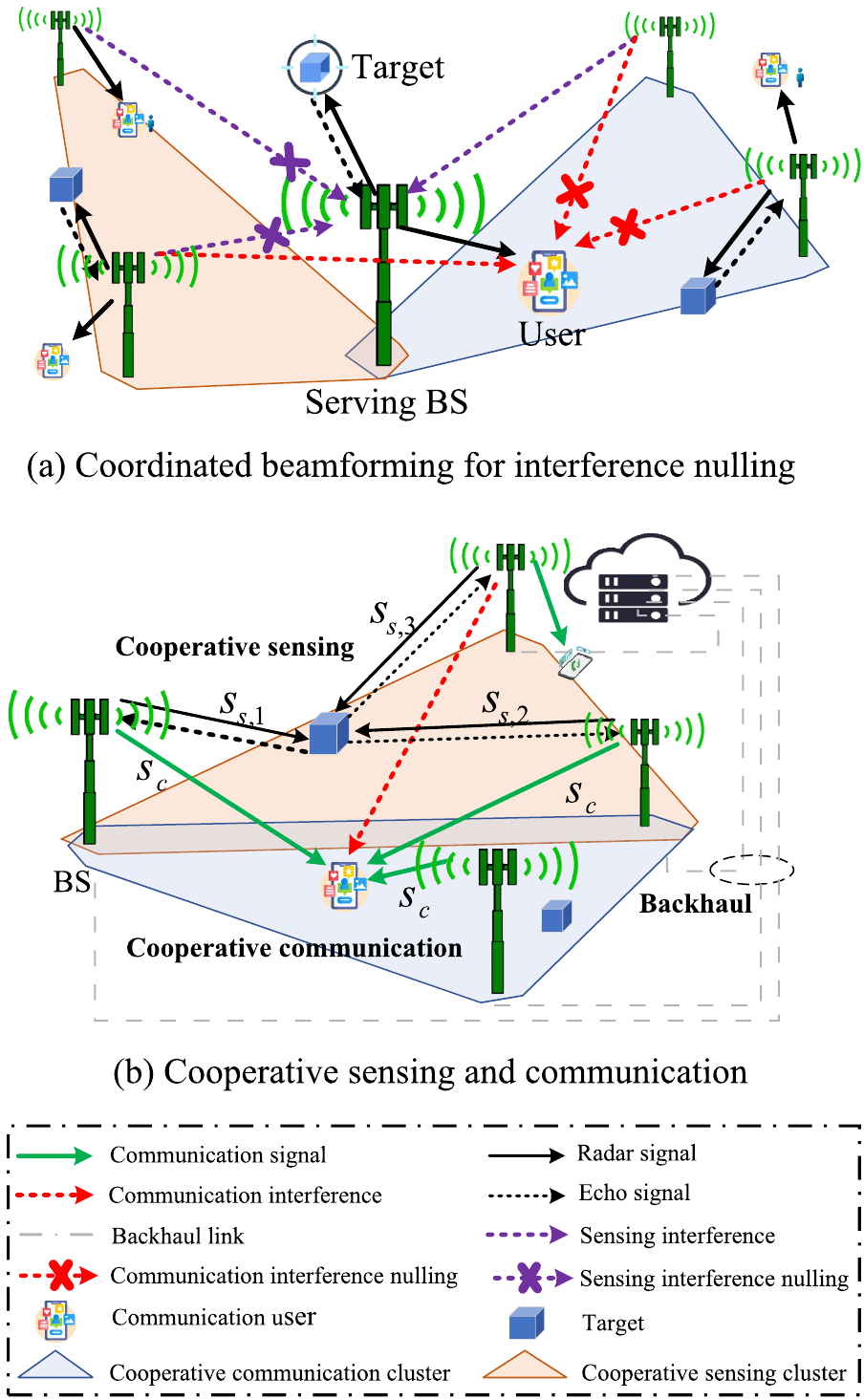}
	\caption{Interference management and cooperative ISAC.}
	\label{figure3}
\end{figure}

\subsection{Basis for ISAC Network Performance Analysis}
To evaluate the end-to-end performance of a dual‐functional sensing and communication network, one must capture the spatial randomness of nodes, channels, and interference across the entire region rather than isolated links; SG offers a natural framework by modeling transmitter and receiver locations as point processes \cite{chiu2013stochastic}. This necessitates a paradigm shift from classical communication-only SG to include targets and radar parameters within the variables of the SG model and capture a range of network-level ISAC metrics such as ISAC SINR distributions, coverage probabilities, detection rates, and false‐alarm probabilities can be derived, enabling a unified comparison of multiplexing gain, diversity gain, and interference‐nulling gain under varying network densities and channel conditions.

\subsubsection{Communication Performance Analysis}
In ISAC-enabled cellular systems, interference management via CBF and CoMP transmission plays a central role in determining end-user performance. By jointly designing transmitter beamformers across neighboring base stations (BSs) and allowing for dynamic user association, the instantaneous signal-to-interference-plus-noise ratio (SINR) at an arbitrary user can be analyzed accordingly. Under a SG model in which BS locations follow a point process, e.g., Poisson point process \(\Phi_b\), the network-wide average data rate is obtained by taking the expectation of the instantaneous spectral efficiency over both the small-scale fading and the BS distribution:  
\begin{equation}
R_c \;=\; \mathbb{E}_{\Phi_b,g_i}\bigl[\log_2\bigl(1 + \mathrm{SINR}_\mathrm{com}\bigr)\bigr]\,,
\end{equation}
where $g_i$ denotes the communication channel gain.
A complementary and widely adopted performance metric is the communication coverage probability, also known as the success probability, which characterizes the probability that a user’s SINR exceeds a preset threshold \(\gamma\) needed for reliable decoding:  
\begin{equation}\label{eq:pcov}
P_c \;=\; \Pr\bigl[\mathrm{SINR}_{\mathrm{com}} > \gamma\bigr]\,.  
\end{equation}  
Here, the threshold \(\gamma\) is chosen according to the minimum required capacity. In practice, one can evaluate \(P_c\) in closed-form (or via tight bounds) by averaging the SINR distribution induced by both CBF patterns and the spatial randomness of \(\Phi_b\).

Together, the coverage probability \(P_c\) and the average rate \(R_c\) provide a comprehensive view of communication performance under advanced interference-coordination schemes. In subsequent sections, these metrics will be extended to jointly capture sensing performance, leading to unified ISAC performance measures.

\subsubsection{Sensing Performance Analysis} \label{Net-Sensing}
Next, we consider three localization schemes, angle-of-arrival (AoA), time-of-flight (ToF), and a hybrid AoA/ToF approach, and show how the network's average positioning performance can be characterized via the expected CRLB.

For the AOA estimate at the receiver $j$, the covariance matrix can be given by
		$
		\mathbf{R}_y = \mathbb{E}\left\{\mathbf{y}_j(t)\mathbf{y}_j^H(t)\right\},
		$
		and perform an eigen-decomposition to separate the signal and noise subspaces. The MUSIC algorithm can be applied to this decomposition to generate a pseudospectrum, where the peak corresponds to the estimated AOA.
		By measuring the AOAs of each monostatic link and bi-static link, the target location can be estimated by maximum likelihood estimation (MLE) \cite{li1993maximum}. For the AOA measurement of the bi-static link from the $j$th BS to the target and then to the $i$th BS, we have
\begin{equation}\label{AngleEstimation}
	\hat {\theta}_{i,j} = \tan ^{-1} \frac{{y}_t-{y_i}}{{x}_t-{x_i}}+n^a_{i,j}.
\end{equation}
In (\ref{AngleEstimation}), $n^a_{i,j}$ denotes the AOA measurement error.  Here, $G_t$ is the transmit beamforming gain, and $\gamma_0$ represents the channel power at the reference distance of 1 m. Then, we transform $N^2$ AOA measurement links into the target location. 
Then, the Fisher information matrix (FIM) of estimating the parameter vector ${{\bm{\psi}}}_t$ for the AOA-based MIMO radar considered is equal to
\begin{equation}\label{FIMexpression_1}
		{\bf{F}}_{\rm{A}} \!  =\! |\zeta_a |^2 \!\sum\nolimits_{j = 1}^N \! {\sum\nolimits_{i = 1}^N \! {\frac{{{{\cos }^2}{\theta _i}}}{{d_j^2 d_i^2}} \! \left[\! {\begin{array}{*{20}{c}}
						{\!\frac{{{{\sin }^2}{\theta _i}}}{{d_i^2}}}&\!{ - \frac{{\sin {\theta _i}\cos {\theta _i}}}{{d_i^2}}}\\
						\!{ - \frac{{\sin {\theta _i}\cos {\theta _i}}}{{d_i^2}}}&\!{\frac{{{{\cos }^2}{\theta _i}}}{{d_i^2}}}
				\end{array}} \!\right]} } \!,
\end{equation}
where  $|\zeta_a |^2$ denotes the common gain \cite{Liu2022SurveyFundamental}, $d_i$ represents the distance from the BS $i$ to the target.
Given the random location of ISAC BSs, the expected CRLB for any unbiased estimator of the target position is given by
\vspace{0mm}
\begin{equation}\label{CRLBAgnleExpression}	
	\mathrm{CRLB}_{\rm{A}}= {\rm{E}}_{\Phi_b, G_t} \left[\operatorname{tr}\left( {\mathbf{F}}_{\mathrm{A}}^{-1}\right)\right],
	\vspace{0mm}
\end{equation}
where $G_t$ denotes the sensing beam gain. In (\ref{CRLBAgnleExpression}), the expectation operation accounts for the randomness in the locations of sensing BSs and the variability in beam power caused by user channel fluctuations, thereby representing the average sensing performance bound across the entire network.

For TOF measurement based localization, from transmitter $j$ to the target and then to receiver $i$, the measured distance between the \( j \)th transmitter and the \( i \)th receiver is given by
\begin{equation}
	\begin{aligned}
		&\hat d_{ij} (\tau_{ij}) = \\
		& \sqrt{\!\left(x_i - x_t\right)^2 + \left(y_i - y_t\right)^2} +  \sqrt{\left(x_j - x_t\right)^2 + \left(y_j - y_t\right)^2} + n^r_{ij},
	\end{aligned}
\end{equation}
where $\quad n^r_{i,j} \sim \mathcal{N}\left(0, \eta_{i,j}^2\right)$. For the TOF-based range estimation \(\hat{d}_{i,j}\), the matched filtering can be applied to the received signal to correlate it with a replica of the transmitted waveform. This process highlights peaks corresponding to time delays caused by targets, which are then converted into range estimates using the speed of light.

Then, the FIM of estimating the parameter vector ${{\bm{\psi}}}_t$ for the TOF measurement radar considered is equal to \cite{sadeghi2021target}
\begin{equation}\label{FIMexpression}
		{\bf{F}}_{\mathrm{R}} = |\zeta_r |^2 \sum\nolimits_{i = 1}^N \sum\nolimits_{j = 1}^N {{d}}_i^{ - \beta }{{{d}}}_j^{ - \beta } {  {\left[ {\begin{array}{*{20}{c}}{a_{ij}^2}&{{a_{ij}}{b_{ij}}}\\
						{{a_{ij}}{b_{ij}}}&{b_{ij}^2}
				\end{array}} \right]} } ,
\end{equation}
where ${a_{ij}} = \cos {\theta _i} + \cos {\theta _j}$, ${b_{ij}} = \sin {\theta _i} + \sin {\theta _j}$, and $|\zeta_r |$ is the common system gain term.
Given the random location of ISAC BSs, the expected CRLB for any unbiased estimator of the target position is given by
\vspace{0mm}
\begin{equation}	
	\mathrm{CRLB}_{\rm{R}} = {\rm{E}}_{\Phi_b, G_t} \left[\operatorname{tr}\left( {\mathbf{F}}_{\rm{R}}^{-1}\right)\right].
	\vspace{0mm}
\end{equation}

Incorporating both AOA and TOF measurements, rather than relying solely on one type of AOA or TOF measurement, can significantly enhance the accuracy and reliability of the localization system, namely the hybrid localization method. It is assumed that the AOA estimation errors and the TOF estimation errors are uncorrelated. Using both AOA and TOF measurements, the expected CRLB for any unbiased estimator of the target position is given by
\begin{equation}\label{LowerCRB}
	\mathrm{CRLB}_{\rm{H}} = {\rm{E}}_{\Phi_b, G_t} \left[\operatorname{tr}\left( \left({\bf{F}}_{\mathrm{A}} +  {\bf{F}}_{\mathrm{R}}\right)^{-1} \right)\right].
\end{equation}

\subsection{Interfernece Management}
BSs in the cooperative ISAC cluster exchange only CSI over dedicated backhaul links, enabling interference‐nulling beamformers, such as zero‐forcing or MMSE, to steer sensing beams into the null spaces of both user and sensing‐receiver channels (see Fig.~\ref{figure3}(c)).  By exploiting extra spatial dimensions at each BS, this approach creates spatial nulls toward selected out‐of‐cell ISAC nodes or users, effectively suppressing dominant inter‐cell interference without sharing any user or target data.  Of course, achieving such nulling requires a surplus of antennas at the BS.  

It was shown in \cite{10735119} that interference nulling can significantly boost both average communication rate and sensing accuracy.  In fact, if one simply maximizes area spectral efficiency (ASE), defined as the number of served users times their average throughput, the optimization will typically favor using all spatial degrees of freedom for multiplexing and diversity, foregoing nulling altogether.  In contrast, when optimizing for sensing, the solution shifts toward allocating spatial resources to cancel BS‐to‐BS interference, since echo reception becomes interference‐limited in antenna‐rich deployments.  

To capture the full trade‐off between communication ASE and sensing ASE under a spatial‐DoF constraint, \cite{10735119} introduces the S–C ASE region, which we now define formally.
\begin{equation}
	\begin{aligned}
			\mathcal{C}_{\mathrm{c}-\mathrm{s}}(K,L,J,Q) \triangleq & \big\{ (\hat r_c, \hat r_s): \hat r_c \leq  T^{\rm{ASE}}_c,  \hat r_s \leq  T^{\rm{ASE}}_s,\\
		&   KL +J(Q-1) \le M_{\mathrm{t}}, J \le J_{\max} \big\},
	\end{aligned}
\end{equation}
where $(\hat r_c, \hat r_s)$ represents an achievable S\&C performance pair. $Q$ and $L$ respectively denotes the number of BSs involved for CBF of sensing and communication, and $J$ and $K$  respectively denotes the number of targets and users served in each cell.

\begin{figure}[t!]
	\centering
	\subfigure[Interference management.]{
		\label{figure5a}
		\includegraphics[width=7.2cm]{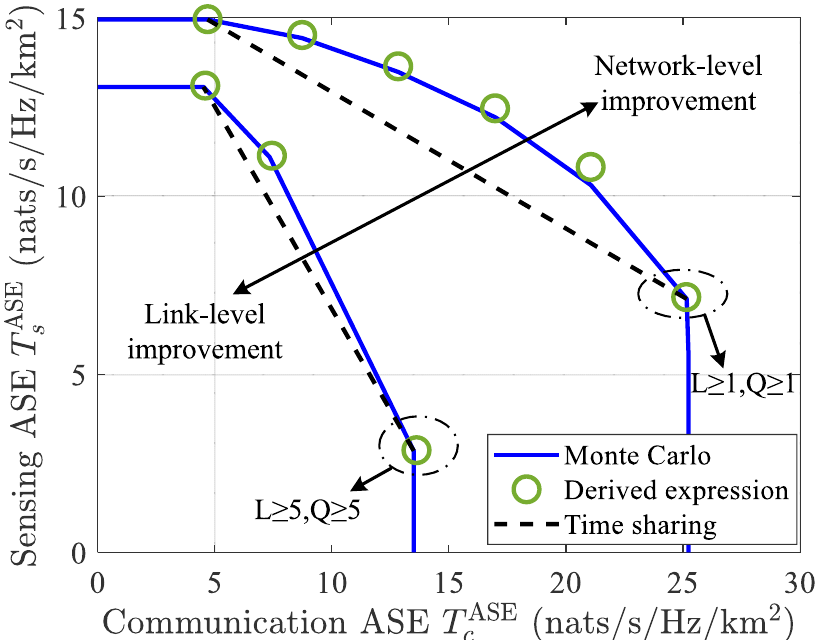}
	}
	\subfigure[Cooperative ISAC.]{
		\label{figure5b}
		\includegraphics[width=7.2cm]{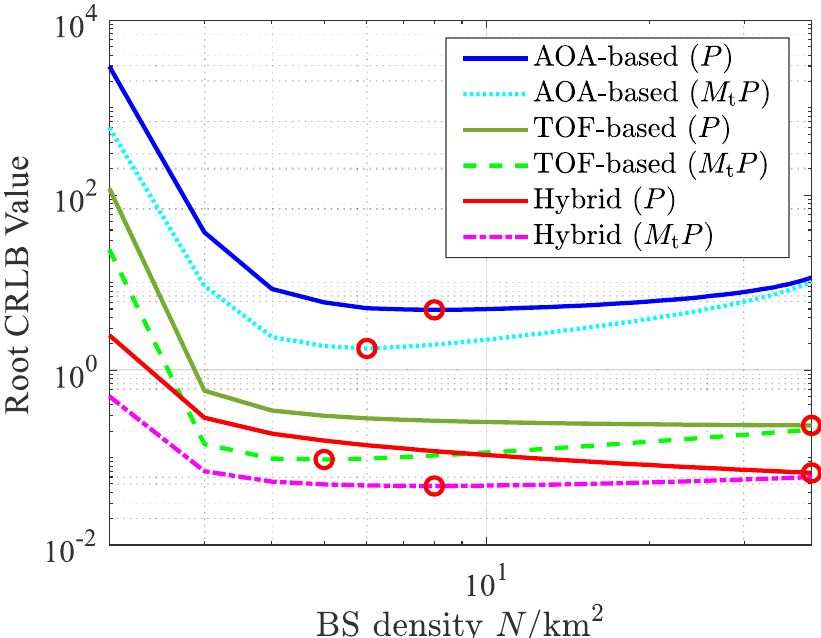}
	}
	\label{figure5ab}
    \caption{Sensing and communication performance for interference management and cooperative ISAC.}
\end{figure}

Moreover, another metric of ISAC networks, i.e., the sum ASE, is defined as a function of the S\&C rates and given by
\begin{equation}\label{WeightedSum}
	T^{\rm{ASE}} = \rho T^{\rm{ASE}}_c + \left(1-\rho \right)T^{\rm{ASE}}_s,
\end{equation}
which represents the total ASE of the ISAC network. Similarly, we can optimize the total ASE $T^{\rm{ASE}}$ to illustrate that the spectral efficiency of ISAC networks is effectively improved compared to sensing-only and communication-only networks. In (\ref{WeightedSum}), $\rho$ represents the weighting factor of S\&C performance. The problem formulation can be expressed as
\begin{alignat}{2}
	\label{P1}
	(\rm{P1}): & \begin{array}{*{20}{c}}
		\mathop {\max }\limits_{J,K,Q,L} \quad  T^{\rm{ASE}}
	\end{array} & \\ 
	\mbox{s.t.}\quad
	& KL +J(Q-1) \le M_{\mathrm{t}}, & \tag{\ref{P1}a}\\
	& J,K,Q,L \ge 1, & \tag{\ref{P1}b} \\
	& J \le J_{\max}. & \tag{\ref{P1}c}
\end{alignat}
It is not difficult to prove that the S-C performance boundary is a convex region, then the optimal total ASE can be obtained by searching the ASE of the boundary point.

Fig. \ref{figure5a} illustrates how imposing minimum cluster‐size requirements, \(L\ge L^{\rm th}\) and \(Q\ge Q^{\rm th}\), to guarantee S\&C quality constrains the achievable network performance envelope.  As these thresholds are raised, the feasible region for the optimization variables \(L\) and \(Q\) shrinks, so the maximum area spectral efficiency that can be supported falls accordingly.  Moreover, the relative benefit of our cooperative design over a simple time‐sharing baseline steadily declines under tighter link‐quality demands: our scheme’s extra throughput comes from selectively relaxing S\&C quality on some links to boost overall scheduling efficiency, but stricter \(L^{\rm th}\) and \(Q^{\rm th}\) preclude this trade‐off and thus cap the collaboration gain.

\subsection{Multi-point Transmission and Multi-static Radar}

Signal-level cooperation in ISAC networks can be implemented in either coherent or non-coherent modes for both communication and sensing.  For example, the scheme in \cite{Meng2024CooperativeTWC} (Fig. 3(d)) works in a non-coherent manner, where the sizes of CoMP and sensing clusters are dynamically selected to maximize joint performance under realistic backhaul constraints.  A complementary analysis in \cite{Meng2024CooperativeTWC} shows that the network-average CRLB scales as \(\mathcal{O}(\ln^2 N)\) with \(N\) cooperating transceivers, improving accuracy but falling short of the \(N^2\) geometric gain of an ideal distributed radar due to path-loss from distant nodes.  
Finally, because each BS must relay both user data and echo measurements over a shared, capacity‐limited backhaul link, enlarging the cooperative cluster size entails a backhaul-load penalty.  This creates an inherent trade-off: larger clusters boost S\&C performance but increase backhaul traffic, requiring careful cluster‐dimensioning in practical ISAC deployments.
\begin{equation}
	R_c + e \times N \le C_{\text{backhaul}},
\end{equation}
where $R_c$ is the average data rate, The variable $e$ represents the data rate required for the sensing cooperation, i.e. for sending pre-processed results (i.e., auto-correlation of the signal $\{s^s_i\}_{i=1}^N$ transmitted by each BS), and $C_{\text{backhaul}}$ denotes the backhaul capacity limitation. Since $N$ BSs are coordinated for positioning, each BS should apply matched filtering to $N$ sensing data signals $\{s^s_i\}^N_{i=1}$, resulting in a total amount of processed data given by $N \times e$.

To strike a flexible tradeoff between the S\&C performance at the network level, we aim for optimizing the cooperative cluster sizes of S\&C under the above backhaul capacity constraints. Our approach focuses on an analytical method for determining the optimal cooperative cluster sizes. This makes it more practically appealing compared to existing clustering schemes employing iterative algorithms, which are more complex and computationally heavy. The optimal cluster parameter can be readily calculated offline from the prior knowledge of the BS load, the ratio of BSs to user densities, and path loss exponents.

\begin{table}[]
	\centering
	\footnotesize
	\caption{Localization Method comparisons \cite{meng2024network}.} 
	\label{Table2}
	\begin{tabular}{
			>{\columncolor[HTML]{E6CFE6}}l 
			>{\columncolor[HTML]{EFEFEF}}l 
			>{\columncolor[HTML]{E6CFE6}}l 
			>{\columncolor[HTML]{EFEFEF}}l }
		\hline
		{\color[HTML]{333333} \textbf{\begin{tabular}[c]{@{}l@{}} Localization\\ Method\end{tabular}}} & {\color[HTML]{333333} \textbf{\begin{tabular}[c]{@{}l@{}}Time \\ Synchronization\end{tabular}}} & {\color[HTML]{333333} \textbf{\begin{tabular}[c]{@{}l@{}}Accuracy \\ factors\end{tabular}}}                                               & {\color[HTML]{333333} \textbf{\textbf{\begin{tabular}[c]{@{}l@{}}CRLB  \\ Scaling law$^\star$\end{tabular}}}} \\  \hline
		{\color[HTML]{333333} \begin{tabular}[c]{@{}l@{}}Angle\\ measurement\end{tabular}}            & {\color[HTML]{333333} No}                                                                      & {\color[HTML]{333333} \begin{tabular}[c]{@{}l@{}}Array aperture size \\ 
				Array orientation\end{tabular}} & {\color[HTML]{333333} $\frac{1}{{\ln N}}$} \\
		\hline 
		{\color[HTML]{333333} \begin{tabular}[c]{@{}l@{}}Range\\ measurement\end{tabular}}            & {\color[HTML]{333333} Yes}                                                                     & {\color[HTML]{333333} Bandwidth}                                                    & { \color[HTML]{333333} $\frac{1}{\ln^2 N}$}        \\
		\hline
		{\color[HTML]{333333} \begin{tabular}[c]{@{}l@{}}Hybrid\\ measurement\end{tabular}}            & {\color[HTML]{333333} Yes}                                                                     & {\color[HTML]{333333} All above}                                                   & { \color[HTML]{333333} $\frac{1}{a \ln^2 N + b \ln N}$}        \\ \hline
		\noalign{\vskip 2mm}   
		\hline
	\end{tabular}
	\begin{tablenotes}
		\footnotesize
		\item $^\star$ $N$ denotes the number of BSs in the cooperative sensing cluster under PPP distribution. $a$ and $b$ denote the parameters of TOF measurements and AOA measurements, respectively.
	\end{tablenotes}
\end{table}

Regarding sensing performance, \cite{meng2024network} investigates three localization methods: AOA-based, TOF-based, and a hybrid approach combining both AOA and TOF measurements, for critically appraising their effects on ISAC network performance, as summarized in Table \ref{Table2}. Our analysis shows that in networks having \( N \) ISAC nodes following a Poisson point process, the localization accuracy of TOF-based methods follows a \( \ln^2 N \) scaling law (explicitly, the Cramér-Rao lower bound (CRLB) reduces with \( \ln^2 N \)). The AOA-based methods follow a \( \ln N \) scaling law, while the hybrid methods scale as \( a\ln^2 N + b\ln N \), where \( a \) and \( b \) represent parameters related to TOF and AOA measurements, respectively. The difference between these scaling laws arises from the distinct ways in which measurement results are converted into the target location. Specifically, when converting AOA measurements to the target location, the localization error introduced during this conversion is inversely proportional to the distance between the BS and the target, leading to a more significant reduction in accuracy as the number of transceivers increases. In contrast, TOF-based localization avoids such distance-dependent errors in the conversion process.

Fig. \ref{figure5b} evaluates localization accuracy with transmit and receive antenna densities fixed at \(\lambda_t=\lambda_r=50\,\mathrm{km}^{-2}\), noise power \(\sigma_s^2=-100\) dB, and bandwidth \(B=10\) MHz. In the legend, “\(P\)” denotes a per–base-station power cap, so each BS radiates the same total power irrespective of its antenna count, whereas “\(M_t\!\cdot\!P\)” scales each BS’s total transmit power in proportion to its number of antennas, keeping the network’s aggregate power constant over the cooperation area.
Under the fixed-\(P\) constraint, both the TOF-only and hybrid AoA/ToF schemes achieve their lowest average CRLB when antennas are fully distributed across all BSs. In contrast, the AoA-only method attains optimal performance by concentrating roughly eight BSs per km\(^2\), thereby exploiting coherent phase combining. When power scales with \(M_t\), the TOF and hybrid approaches shift to a hybrid deployment, combining centralized arrays for beamforming gain with distributed nodes for macro-diversity, striking a balance that neither extreme alone can achieve. At their respective optima, the hybrid scheme shrinks the localization error to just 1.3\% of the AoA-only error and 28.8\% of the TOF-only error, underscoring its robustness across different backhaul power models.

\section{Cooperative Signaling Design for Coherent and Non-Coherent Distributed ISAC} \label{sec::Signaling}
This section addresses the signaling design considerations for enabling cooperative sensing and communication functionalities in distributed ISAC systems. In contrast to ISAC signaling design in collocated single-node systems \cite{liu2021cramer, liu2020joint, hua2023optimal}, where sensing performance is primarily driven by spatial beamforming of radar waveforms, the target localization accuracy in distributed MIMO sensing depends on both TOF and AOA estimation performance. Similarly, the performance of cooperative communication is influenced by the signaling strategies employed across distributed ISAC nodes, such as CBF and joint transmission CoMP. As a result, transmit signaling design in distributed ISAC systems must carefully balance sensing and communication requirements, while leveraging the unique opportunities and addressing the inherent challenges posed by the spatial separation of transceivers.

\subsection{Communication Performance Metric for Signaling Design}
As a key performance metric for cooperative communication design, we introduce the signal-to-interference-plus-noise ratio (SINR) at the typical user. SINR has been widely adopted in existing cooperative ISAC signaling designs and can be further extended to evaluate other metrics such as spectral efficiency or sum-rate. Notably, it reveals that SINR metric has different formation depending on cooperation scheme and phase-coherency among ISAC nodes. Therefore, the proper selection of the metric yields the signaling design along different system-level coordination and cooperation schemes. 

\subsubsection{Coordinated beamforming (CBF)}
For instance, in CBF mode, where inter-node interference is treated as a harmful signal to the typical user, as shown in \eqref{comm_coordinated}, the resulting communication SINR is affected by multi-user interference, inter-node interference, and interference from sensing signals. Although this approach is suboptimal compared to cooperative communication via joint transmission, it has been widely adopted in cooperative signaling designs \cite{yang2024coordinated, cheng2024optimal, babu2024precoding, cheng2025networked, jiang2025federated, wei2023integrated} due to its lower complexity in inter-node information sharing and reduced requirements for time-frequency synchronization. The SINR of user $k$ served by ISAC node $1$ for CBF is expressed as
\begin{equation}
    \gamma_{\mathrm{CB},k} = \frac{|\tilde{\mathbf{h}}^{H}_{k,1} \mathbf{w}^{c}_{k,1}|^2}{\sum_{n = 1}^{N} |\tilde{\mathbf{h}}^{H}_{k,n} \mathbf{X}_{n}|^2 - |\tilde{\mathbf{h}}^{H}_{k,1} \mathbf{w}^{c}_{k,1}|^2  + \sigma_{c}^2 },
    \label{Eqn::SINR_CB}
\end{equation}
where $\sigma_c^2$ is the noise power at the communication user receiver. It can be observed that, in CBF, all signals except for the dedicated data signal from node 1 act as interfering signals at the typical user \cite{cheng2024optimal}. It is worth noting that while CBF inherently avoids the need for transmit data sharing among ISAC transmitters, it is not entirely free from data exchange requirements in the perspective of sensing. Specifically, the transmitted signals may need to be shared with designated sensing receivers for demodulating sensing target channels or collecting raw data at a CPU, potentially introducing additional data sharing overhead.

\subsubsection{Non-Coherent Joint Transmission}
A non-coherent joint transmission scheme can be considered, which does not require phase-level synchronization among ISAC nodes. As a result, it is robust to phase mismatches and hardware imperfections across distributed nodes. In this approach, the useful communication signals transmitted by cooperative ISAC nodes are non-coherently combined at the user receiver, corresponding to a power combining of multiple transmitted signals \cite{tanbourgi2014tractable, vu2020noncoherent, wang2024joint, d2024coherent, han2025signaling}. The resulting SINR for user $k$ under non-coherent joint transmission is expressed as
\begin{equation}
    \gamma_{\mathrm{NCJT},k} = \frac{\sum_{n=1}^{N} |\tilde{\mathbf{h}}^{H}_{k,n} \mathbf{w}^{c}_{k,n}|^2}{\sum_{n=1}^{N} \left( |\tilde{\mathbf{h}}^{H}_{k,n} \mathbf{X}_{n}|^2 - |\tilde{\mathbf{h}}^{H}_{k,n} \mathbf{w}^{c}_{k,n}|^2 \right) + \sigma_{c}^2},
    \label{Eqn::SINR_NCJT}
\end{equation}
It can be observed that the transmitted sensing signals and multi-user interference from all ISAC nodes are non-coherently combined at the receiver, resulting in an increased interference power.

\subsubsection{Coherent Joint Transmission}
The coherent joint transmission CoMP scheme, where multiple nodes collaboratively serve a common user, leverages signal combining gain by achieving phase-level synchronization across the ISAC network. Although acquiring tight synchronization is challenging, this cooperative approach offers the optimal cooperation gain for communication performance. Coherent joint transmission is widely adopted in cell-free ISAC systems and in scenarios involving multiple base station cooperation rather than mere coordination \cite{zhang2025coordinated, liu2024joint, mao2024communication, demirhan2024cell, liu2024cooperative, xu2025distributed}. For coherent joint transmission, the SINR for user $k$ is given by
\begin{equation}
    \gamma_{\mathrm{CJT},k} = \frac{|\sum_{n=1}^{N}\tilde{\mathbf{h}}^{H}_{k,n} \mathbf{w}^{c}_{k,n}|^2}{|\sum_{n=1}^{N}\tilde{\mathbf{h}}^{H}_{k,n} \mathbf{X}_{n}|^2 - |\sum_{n=1}^{N}\tilde{\mathbf{h}}^{H}_{k,n} \mathbf{w}^{c}_{k,n}|^2 + \sigma_{c}^2},
    \label{Eqn::SINR_CJT}
\end{equation}
While coherent cooperation can theoretically achieve optimal performance under perfect phase synchronization, practical implementation remains challenging. In widely separated nodes, the CSI available at the transmitters $\hat{\mathbf{h}}^{H}_{k,n} = e^{-j \phi_{n}} \tilde{\mathbf{h}}^{H}_{k,n}$ may differ from the actual CSI $\tilde{\mathbf{h}}^{H}_{k,n}$ due to hardware impairments such as phase noise. Consequently, the SINR expression in \eqref{Eqn::SINR_CJT} represents an upper-bound performance achievable under the assumption of perfect CSI and ideal synchronization across the ISAC network.

\subsection{Sensing Performance Metric for Signaling Design}
For distributed ISAC signaling design, the sensing metric must be carefully selected to leverage the unique opportunities offered by distributed radar, enhancing target detection and localization performance through increased spatial diversity. While the target detection probability \cite{cheng2024optimal, ren2024secure, Igno-li2025distributed} and the SINR for the target \cite{liu2024joint, demirhan2024cell} can also serve as a sensing performance metric in cooperative signaling design for distributed ISAC, this section introduces the target localization accuracy of coherent and non-coherent MIMO sensing.

\subsubsection{Localization CRLB for Non-Coherent MIMO Sensing}
For the non-coherent case, one can leverage TOF and AOA information to localize targets based on individual node measurements, similar to the hybrid localization approach discussed in Section \ref{Net-Sensing}. Let $\mathbf{\Psi}$ represent the set of all real-valued unknown parameters, given by $\mathbf{\Psi} = [\boldsymbol{\theta}^T, \boldsymbol{\tau}^T, \mathbf{b}_{\textsc{R}}^T, \mathbf{b}_{\textsc{I}}^T]^T \in \mathbb{R}^{(2UN+2UN^2) \times 1}$, where $\boldsymbol{\tau}$ and $\boldsymbol{\theta}$ are the collection of TOF and AOA parameters, and $\mathbf{b} = [\beta_{1,1}^1, \beta_{1,1}^2, \ldots, \beta_{n,i}^u, \ldots, \beta_{N,N}^U]^T \in \mathbb{C}^{UN^2 \times 1}$ with $\mathbf{b}_{\textsc{R}} = \text{Re}(\mathbf{b})$ and $\mathbf{b}_{\textsc{I}} = \text{Im}(\mathbf{b})$.
Since both $\boldsymbol{\theta}$ and $\boldsymbol{\tau}$ contribute to estimating target positions in Cartesian coordinates, we alternatively define the parameter vector as
$\boldsymbol{\Theta}_{nc} = [\mathbf{x}^T, \mathbf{y}^T, \mathbf{b}_{\textsc{R}}^T, \mathbf{b}_{\textsc{I}}^T]^T \in \mathbb{R}^{(2UN^2 + 2U) \times 1},$ where $\mathbf{x} = [x_1, x_2, \ldots, x_U]^T \in \mathbb{R}^{K \times 1}$ and $\mathbf{y} = [y_1, y_2, \ldots, y_U]^T \in \mathbb{R}^{K \times 1}$. Thus, the CRLB for target localization in the non-coherent case is obtained from the FIM with respect to $\boldsymbol{\Theta}_{nc}$, expressed as
\begin{equation}\label{SignalingCRB_non}
	\mathrm{CRLB}_{\rm{nc}} =  \operatorname{tr} \left({\bf{F}}^{-1}_{\rm{nc}}(\boldsymbol{\Theta}_{\rm{nc}}) \right).
\end{equation}
It is noted that the complex amplitudes $\beta_{n,m}^u$ act as nuisance parameters and do not directly affect target localization in non-coherent distributed MIMO radar sensing. A detailed derivation of the CRLB can be found in \cite{han2025signaling, godrich2010target}.

\subsubsection{Localization CRLB for Coherent MIMO Sensing}
The coherent processing of distributed ISAC systems assumes phase synchronization among ISAC nodes, enabling the exploitation of additional delay-dependent phase information embedded in the target complex amplitude $\beta_{n,m}^u$ \cite{yang2011phase}. To derive the CRLB for coherent localization, the complex amplitude term is appropriately reformulated. Assuming the signal bandwidth is much smaller than the carrier frequency $f_c$, the complex amplitude can be approximated as
\begin{equation}\label{Eqn::19}
    \beta_{n,m}^u = |\beta_{n,m}^u| e^{-2\pi f_c \tau_{n,m}^u},
\end{equation}
where $|\beta_{n,m}^u|$ denotes the average magnitude of the reflection from target $u$. In this case, the TOF information embedded in the phase of $\beta_{n,m}^u$ contributes to localization, providing a coherent gain proportional to $f_c$. Accordingly, the FIM differs from the non-coherent case, where the unknown parameters are given by $\boldsymbol{\Theta}_{cc} = [\mathbf{x}^T, \mathbf{y}^T, \mathbf{b}_{\textsc{cc}}^T]^T \in \mathbb{R}^{(UN^2 + 2U) \times 1}$, with $\mathbf{b}_{\textsc{cc}} = [|\beta_{1,1}^1|, |\beta_{1,1}^2|, \ldots, |\beta_{n,i}^u|, \ldots, |\beta_{N,N}^U|]^T$. Thus, the CRLB for coherent localization is given by
\begin{equation}\label{SignalingCRB_co}
	\mathrm{CRLB}_{\rm{cc}} =  \operatorname{tr} \left({\bf F}^{-1}_{\rm{cc}}({\boldsymbol \Theta_{\rm{cc}}}) \right).
\end{equation}
The detailed derivation of the coherent CRLB can be found in \cite{sadeghi2021target, godrich2010target}.

\subsection{S\&C Trade-off Design in Distributed ISAC}
Building on the communication and sensing metrics, distributed ISAC signaling can be designed by formulating a conventional CRB minimization problem under communication SINR constraints \cite{liu2021cramer, han2025signaling}. This can be generally expressed as the following optimization problem under a per-antenna power constraint $P_{T}$:
\begin{subequations}\label{Eqn::P1-0}
    \begin{align}
    & \underset{\{\mathbf{X}_{n}\}_{n=1,2,\ldots,N}}{\text{minimize}}
    & & \text{tr}\left([\mathbf{F}(\boldsymbol{\Theta})]^{-1}\right) \label{Eqn::P1-0a} \\
    & \text{subject to}
    & & \left[\mathbf{R}_n\right]_{m,m} \leq P_{T}, \quad \forall n, m, \label{Eqn::P1-0b} \\
    & & & \gamma_{k} \geq \Gamma_{c}, \quad \forall k, \label{Eqn::P1-0c}
    \end{align}
\end{subequations}
where $\mathbf{R}_n = \mathbb{E}[\mathbf{X}_n \mathbf{X}_n^H]$ denotes the transmit covariance matrix at node $n$, and $m$ indexes the antennas at each node. The SINR for each user must satisfy the threshold $\Gamma_{c}$, ensuring that sensing operations do not compromise the communication quality of service (QoS). The communication metric can also be readily modified from an SINR constraint to a sum-rate constraint, depending on system requirements. Furthermore, the problem can alternatively be formulated as a weighted-sum design of sensing and communication performance, as discussed in \cite{babu2024precoding}. The formulated signaling design problem can be addressed under various signaling conditions, which will be discussed in the following section.

\subsubsection{Beamforming Optimization}
With the CRB-SINR trade-off designs, several works have been reported on beamforming optimization for distributed ISAC signaling \cite{gao2023cooperative, yang2024coordinated, zhang2025coordinated, babu2024precoding}. These solutions typically exploit optimization techniques such as semi-definite relaxation (SDR) relaxing the rank-one constraint on beamforming matrices and successive convex approximation (SCA) to overcome the non-convexity of the original problem formulation. Notably, the works in \cite{yang2024coordinated, zhang2025coordinated} extend conventional beamforming designs by introducing time-asynchronous errors into the target localization CRB in the formulated problem \eqref{Eqn::P1-0}. By constructing a hybrid CRB that incorporates the prior statistics of timing errors, they offer robust signaling designs capable of mitigating the impact of imperfect time synchronization across distributed ISAC nodes. Such approaches are crucial for practical implementations where tight network-wide synchronization is difficult to maintain.

\subsubsection{Joint TOF-AOA Optimization}
The study in \cite{han2025signaling} presented a signaling strategy aimed at improving target localization accuracy through the optimization of per-subcarrier transmissions, altering the communication constraint in \eqref{Eqn::P1-0}. It enables the joint estimation of TOF and AOA parameters. Focusing on OFDM-based distributed ISAC systems,  it demonstrates that optimal target localization necessitates a dedicated per-subcarrier signal design, . In addition, the work addresses the complexity challenges posed by per-subcarrier optimization by introducing methods to maintain signal orthogonality across multiple ISAC nodes, effectively reducing the design burden. Three signaling schemes are introduced: an optimal design that delivers the highest localization accuracy at the cost of significant computational complexity, an orthogonal design that simplifies implementation by enforcing inter-node signal orthogonality, and a beamforming-only design that further lowers complexity by prioritizing AOA estimation accuracy. These strategies offer practical and flexible solutions for distributed ISAC systems, enabling a favorable balance between sensing performance and implementation complexity.

\subsubsection{Symbol-Level Precoding}
The work in \cite{babu2024precoding} presents a symbol-level precoding strategy for CBF and joint transmission CoMP in a multi-cell setting. Unlike fully distributed sensing systems, it concentrates on bi-static sensing scenarios, where the CRLB associated with problem \eqref{Eqn::P1-0} is adapted to focus the AOA estimation CRB in bi-static configurations. In addition, the conventional communication SINR constraint is modified to match the symbol-level precoding framework, offering improved communication reliability compared to the block-level precoding used in \eqref{Eqn::P1-0}. The resulting optimization problem is tackled using SDR and alternating optimization techniques. It is demonstrated that integrating CoMP-based cooperative transmission or CBF with symbol-level precoding significantly enhances ISAC performance in distributed setups. For more details on symbol-level precoding, readers are referred to \cite{masouros2015exploiting}.  

\section{Over-the-Air Synchronization Techniques for Distributed ISAC} \label{sec::Sync}
Throughout the performance analysis of network-level ISAC, it has been consistently observed that distributed cooperation significantly enhances both communication and sensing capabilities. However, the achievable performance gains are fundamentally constrained by practical system limitations, particularly the difficulty of achieving precise time and frequency synchronization across distributed nodes. These challenges primarily stem from hardware impairments, including clock jitter in transmit/receive triggering and phase noise in signal sources \cite{jagannathan2004effect, de2024sensing}. Such impairments introduce inconsistencies across the ISAC network, undermining the coherence required for effective distributed ISAC operations.

Maintaining synchronization among physically separated nodes via over-the-air transmissions remains highly challenging, largely due to the dependency on reliable line-of-sight (LoS) links and the constraints imposed by limited SNR. For example, global navigation satellite systems (GNSS), such as GPS, can provide timing and frequency synchronization with an accuracy on the order of 100 ns \cite{zou2015network}. Nonetheless, GNSS-based solutions typically perform poorly in indoor or dense urban environments \cite{jungnickel2008synchronization}, and the synchronization precision they offer is insufficient for distributed ISAC systems targeting centimeter-level localization accuracy, requiring picoseconds-level synchronization \cite{merlo2023picosecond, prager2020wireless}. Therefore, achieving finer synchronization remains a critical bottleneck for fully realizing the cooperative gains of networked ISAC. In this section, we focus on over-the-air synchronization methods specifically tailored for distributed ISAC systems, aiming to overcome these practical challenges and enable high-precision, scalable network-level cooperation.

\subsection{Over-the-Air Synchronization: Bi-Static Case}
Although bi-static ISAC systems (with $N=2$ nodes, TX/RX separation) represent a simplified form of distributed sensing systems (with $N>2$), they are also practical and widely applicable in real-world deployments \cite{10443836}. Moreover, many distributed ISAC systems can be modeled as a set of interconnected bi-static links \cite{wxy-1}, implying that synchronization algorithms developed for bi-static systems can provide valuable insights and even serve as baseline solutions for more general distributed asynchronous ISAC systems.

In bi-static configurations, the non-co-located transmitter and receiver operate with independent oscillators, which inevitably introduce carrier frequency offset (CFO) and time offset (TO) into signals due to hardware imperfections, temperature variations, and other factors. The CFO and TO offsets result in velocity and range sensing ambiguities, respectively \cite{ni2021uplink}, significantly degrading the system's sensing accuracy \cite{wu2024sensing}. Specifically, with the TO and CFO between nodes, (\ref{Eqn::8}) can be reformulated as
\begin{align}
\mathbf{Y}_{s,n}^{u}\!=\!\!\sum_{m=1}^{N}\!\!\left( \beta_{n,m}^{u}\!\mathbf{a}_{r,n}(\theta_{n}^{u}) \left(\!\mathbf{a}_{t,m}^T(\theta_{m}^{u}) \mathbf{X}_m\!\mathbf{P}_{\tau_{n,m}^{u}}\!\mathbf{P}_{\Delta_{n,m}^\textrm{t}}\!\mathbf{P}_{\Delta_{n,m}^\textrm{f}}\!\right)\!\right), \label{Eqn::27}
\end{align} 
where $\mathbf{P}_{\Delta_{n,m}^\textrm{t}}$, $\mathbf{P}_{\Delta_{n,m}^\textrm{f}}$, and $\mathbf{P}_{\tau_{n,m}^{u}}$ are diagonal matrices representing the phase shifts caused by the TO $\Delta_{n,m}^\textrm{t}$, the CFO $\Delta_{n,m}^\textrm{f}$ between the $n$th and $m$th nodes, and the round-trip delay $\tau_{n,m}^{u}$, respectively. Particularly, the $l$th element of the diagonal vector of $\mathbf{P}_{\Delta_{n,m}^\textrm{t}}$, $\mathbf{P}_{\Delta_{n,m}^\textrm{f}}$, and $\mathbf{P}_{\tau_{n,m}^{u}}$ are $e^{j\Delta_{n,m}^\textrm{t}\phi_1}$, $e^{j(l-1)\Delta_{n,m}^\textrm{f}\phi_2}$, and $e^{j\tau_{n,m}^{u}\phi_1}$, respectively, where $\phi_1$ and $\phi_2$ are constant parameters. It is important to note that the TO and CFO equal zero when $m=n$. Therefore, for clarity, we only focus on the bi-static passive sensing, where the $1$st node transmits and the $2$nd node receives the signal.
To mitigate the impact of CFO and TO to ensure reliable sensing performance, four primary types of synchronization schemes have been proposed for bi-static systems. 

\subsubsection{Cross-Antenna-Based Synchronization}
The first type extracts CFO and TO by utilizing signal (or CSI) division and cross-correlation across different receive antennas, as exemplified by the CSI-ratio-based scheme \cite{FarSense,zeng2020multisense,10678871} and the cross-antenna cross-correlation (CACC) method \cite{IndoTrack,ni2021uplink}. Both approaches utilize the signal received by a reference antenna as a designated “synchronization reference”. 

To illustrate the core operation of CACC, we formulate $\mathbf{Y}_{s,2}$ in (\ref{Eqn::9}) as 
\begin{align}
\mathbf{Y}_{s,2}=\mathbf{Y}_{s,1}^{\textrm{LOS}}+\sum_{u=1}^{U}(\mathbf{Y}_{s,1}^{u,\textrm{NLOS}})+\mathbf{Z}_{s,1}, 
\end{align}
where $\mathbf{Y}_{s,1}^{\textrm{LOS}}$ and $\mathbf{Y}_{s,1}^{u,\textrm{NLOS}}$ denote the LOS and NLOS signal components, respectively. Then, performing cross-correlation between different rows of $\mathbf{Y}_{s,2}$ yields several terms: the cross-correlation of LOS components, cross-terms between LOS and NLOS, cross-correlation among NLOS components, and noise. Since the static LOS component typically has much higher power than the dynamic NLOS components, the NLOS-NLOS correlation term can be neglected, and the LOS-LOS term appears as a constant. In the LOS-NLOS cross-term, the shared phase shift caused by TO and CFO cancels out the effect of $\mathbf{P}_{\Delta_{n,m}^\textrm{t}}$ and $\mathbf{P}_{\Delta_{n,m}^\textrm{f}}$, allowing the time delay and velocity to be extracted. However, CACC is only applicable in LOS scenarios with multiple antennas.

The CSI-ratio-based scheme leverages the division of CSI across antennas to form a Möbius transform regarding the phase shift induced by moving targets \cite{FarSense}. Due to the properties of Möbius transforms, when the phase shift varies with the target's motion, the transform traces a circular trajectory on the complex plane. The displacement and velocity can then be estimated by analyzing the angular velocity and phase evolution of this trajectory. Thanks to the properties of Möbius transforms, this method is applicable in any scenario with a static path. However, it only supports displacement estimation and does not enable full target localization.

\subsubsection{Path-Resolved Synchronization}
The second type addresses CFO and TO by separating received signals into dynamic and static components using high-angular-resolution antenna arrays or high-complexity path separation algorithm \cite{10207823,10443836,9804861,brunner2024bistatic }. Once the signal paths are separated, TO and CFO can, in principle, be estimated from the static-path signals and subsequently used to compensate the dynamic-path signals. This approach effectively mitigates synchronization errors. However, it requires arrays with very high angular resolution to distinguish between individual paths, which poses significant challenges for practical deployment.

\subsubsection{Fingerprint-Based Synchronization}
\begin{figure}[t!]
	\centering
	{\includegraphics[width=7.2cm]{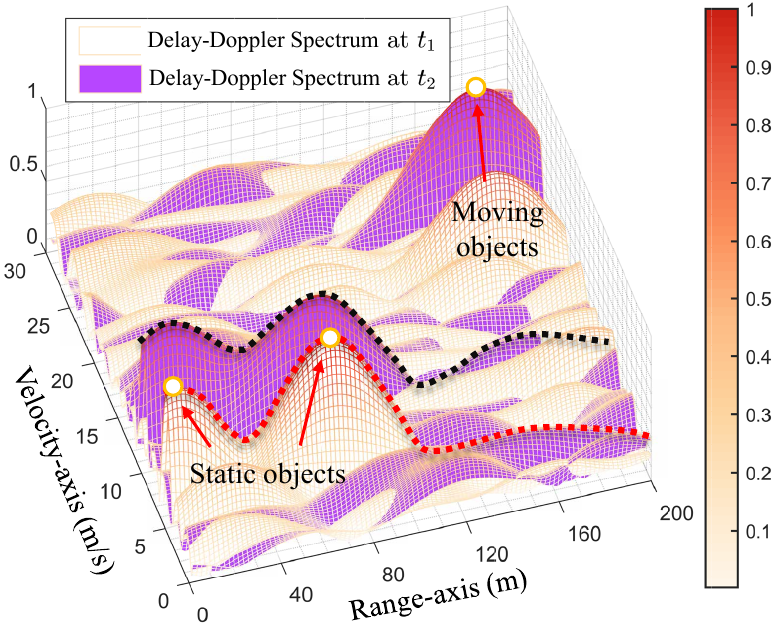}}
        \caption{A fingerprint spectrum realization \cite{wxy}. The red and black lines represent the fingerprint spectra at times $t_1$ and $t_2$, respectively. The peaks corresponding to static objects appear around 5 m/s due to the velocity ambiguity induced by CFO.}
        \label{fig6a}
\end{figure}

The third type estimates CFO and TO by leveraging the delay-Doppler spectrum of signals reflected from static objects, treating it as an environment-specific \textit{fingerprint spectrum} \cite{wxy,10694270}. This spectrum is uniquely determined by the spatial distribution of static objects in a given environment, akin to a fingerprint. An intuitive illustration of such a fingerprint spectrum for a specific environment is shown in Fig. \ref{fig6a}. Notably, the fingerprint spectrum exhibits a two-dimensional \textit{circular shift property} within the delay-Doppler spectrum matrix, where variations in TO and CFO cause circular shifts along the row (delay) and column (Doppler) dimensions, respectively. Based on this property, CFO and TO can be accurately estimated by performing cross-correlation between the fingerprint spectrum vector and the full delay-Doppler spectrum matrix acquired at different time instants. 

\begin{figure}[t!]
        \centering
        {\includegraphics[width=7.2cm]{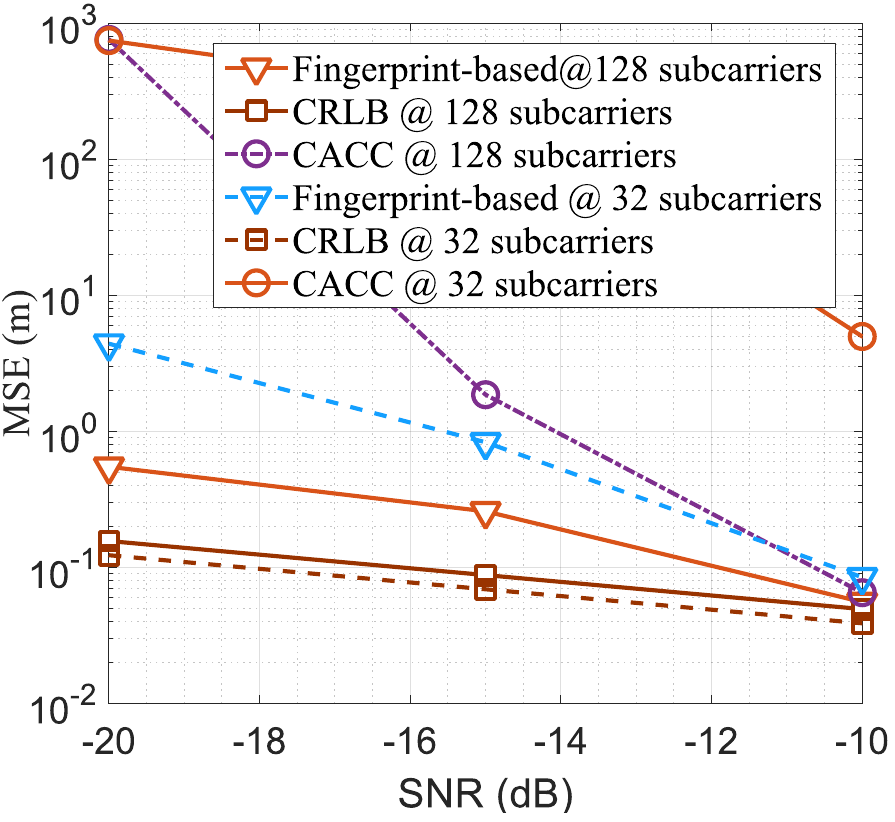}}
        \caption{Range estimation performance comparison between the fingerprint-based scheme and the CACC method under asynchronous ISAC systems with different numbers of subcarriers.}
        \label{fig6b}	
\end{figure}

For clarity, we denote the fingerprint spectrum vector at time $t_1$ as $\boldsymbol{\beta} \in \mathbb{C}^{1 \times G_2}$, and the full delay-Doppler spectrum matrix at time $t_2$ as ${\boldsymbol{\Xi}} \in \mathbb{C}^{G_1 \times G_2}$, where a specific row of ${\boldsymbol{\Xi}}$ represents the fingerprint spectrum at $t_2$. Then, the TO $\Delta_{n,m}^\textrm{t}$ and CFO $\Delta_{n,m}^\textrm{f}$ can be estimated as
\begin{equation}
    \begin{aligned}
    \{&\frac{\Delta_{n,m}^\textrm{f}}{F_{\rm R}},\frac{\Delta_{n,m}^\textrm{t}}{T_{\rm R}}\}=\mathop{\rm max}\limits_{i,q}\Big|\sum_{p=1}^{G_2}\frac{{\boldsymbol{\Xi}}[i,(q+p)\ {\rm mod}\ {G_2}]{\boldsymbol{\beta}^*[p]}}{|{\boldsymbol{\beta}}|^2}\Big|,\\
    &{\rm for}\ i=1,\cdots,G_1\ {\rm and} \ q=1,\cdots,G_2,
    \end{aligned}
\end{equation} 
where $F_{\rm R}$ and $T_{\rm R}$ are the constants related to the number of rows and columns of the delay-Doppler matrix.
This method achieves high-precision CFO and TO estimation under both LOS and NLOS conditions, as well as in single- and multi-target scenarios, demonstrating strong applicability. Fig. \ref{fig6b} compares the performance of the fingerprint spectrum-based synchronization with the CACC scheme in OFDM ISAC systems. However, a key limitation of the fingerprint spectrum approach is the relatively complex process of acquiring accurate fingerprints in practical deployment scenarios.


\subsubsection{Cooperation-Based Synchronization}
This type estimates CFO and TO by exploiting the cross-correlation between active and passive sensing echoes in bi-static scenarios with joint active and passive sensing \cite{10616023,9848431}. The author treats the active sensing echo, whose TO and CFO are zero, as a reference. By comparing it with the passive sensing echo, which suffers from unknown TO and CFO, the algorithm can identify and estimate these offsets based on the phase discrepancies between the two. This scheme is robust in NLOS scenarios. However, to ensure accurate synchronization, the active and passive sensing operations must be conducted over different frequency bands and the receiving power of active sensing echo and passive sensing echo should be close.

\subsection{Over-the-Air Synchronization: Distributed Case}
Extending the concept of bi-static synchronization, it is recognized that the sensing capabilities inherent in ISAC systems provide a novel opportunity to achieve synchronization among asynchronous nodes without a direct LOS link. Unlike conventional bi-static ISAC scenarios, distributed ISAC systems with collocated transceivers (as illustrated in Fig.~\ref{Fig_KW1}(b)) can leverage pair-wise bi-static links formed through reflections from multiple objects or scatterers in the environment. This unique feature offers an additional advantage for distributed ISAC, enabling enhanced synchronization performance by exploiting naturally occurring environmental reflections.

 \begin{figure}[t!]
    \centering
    {\includegraphics[scale=0.55]{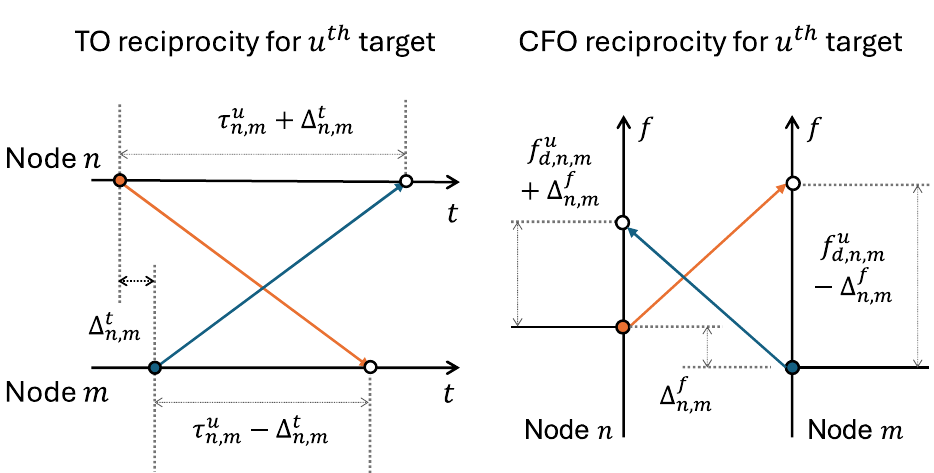}} 
    \caption{Offset reciprocity of reciprocal bi-static links in distributed ISAC systems.}
    \label{f6}
\end{figure}

The approach introduced in \cite{han2025over} exploits the observation that when two nodes independently sense the same objects from different spatial locations, the resulting TO and CFO exhibit a reciprocal relationship. Specifically, these offsets are equal in magnitude but opposite in sign, as illustrated in Fig.~\ref{f6}. This behavior, known as offset reciprocity (OR), serves as the foundation for a high-accuracy synchronization framework. The OR property in distributed ISAC systems can be mathematically expressed as
\begin{align}
    \Delta_{n,m}^t &= -\Delta_{m,n}^t, \\
    \Delta_{n,m}^f &= -\Delta_{m,n}^f,
\end{align}
where $\Delta_{n,m}^t$ and $\Delta_{n,m}^f$ represent the TO and CFO between nodes $n$ and $m$ measured at the receiving node $n$, respectively. By utilizing the OR property, TO and CFO between node pairs can be determined without prior knowledge of either the targets' positions or the ISAC nodes' locations, making the method highly adaptable and scalable for distributed ISAC systems. Furthermore, it enables to efficiently implement super-resolution offset estimation. This technique for distributed ISAC improves synchronization accuracy without relying on dedicated reference signals, providing a practical and efficient solution for large-scale distributed ISAC networks.

\begin{figure}[t!]
    \centering
    {\includegraphics[scale=0.7]{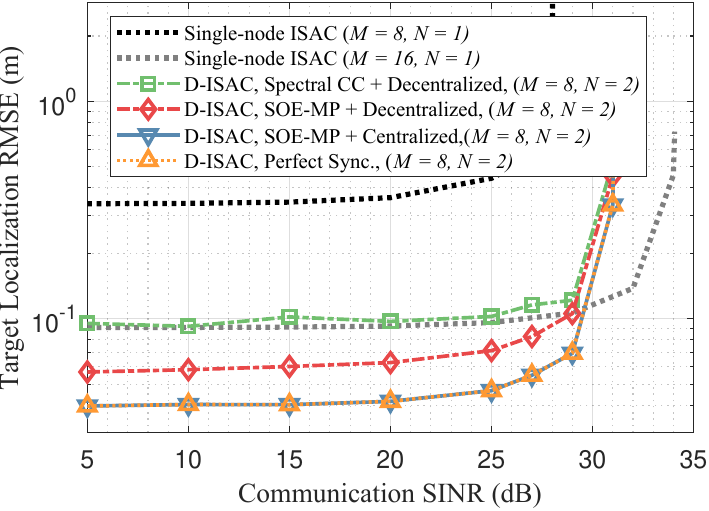}} 
    \caption{Performance gains of distributed ISAC (D-ISAC) systems employing various synchronization methods, compared to a conventional single-node ISAC baseline. $M$ represents the number of transmit antennas at each ISAC node.}
    \label{f7}
\end{figure}

The ISAC trade-off performance resulting from different synchronization techniques supports the analysis, as illustrated in Fig.~\ref{f7}, where spectral cross-correlation (CC) denotes the on-grid offset estimation with FFTs \cite{sigg2024over, aguilar2024uncoupled} and SOE-MP represents the super-resolution offset estimation through matrix pencil method with OR property. It is observed that decentralized processing is highly sensitive to time offsets between ISAC nodes. This is because only a single monostatic link cannot effectively compensate for bistatic measurement errors, leading to degraded localization accuracy. Moreover, distributed ISAC performance relying on synchronization through spectral cross-correlation exhibits a noticeable performance gap compared to the ideal case without synchronization impairments. This underscores the advantage of the super-resolution offset estimation techniques, which enable fine-grained offset estimation and thereby improve synchronization accuracy.

Finally, distributed ISAC enhances target localization by exploiting spatial diversity through distributed MIMO radar architectures but remains vulnerable to synchronization inaccuracies. A comparison between noncoherent distributed ISAC and a single-node collocated ISAC system reveals that, when each node is equipped with the same number of antennas, distributed ISAC achieves superior localization accuracy and SINR through power combining effects. However, when considering a single-node ISAC system with an equivalent total number of antennas across all nodes, the single-node system achieves higher SINR due to coherent combining gain. While distributed ISAC benefits from spatial diversity, synchronization based on spectral cross-correlation (with 250 ps accuracy) limits its ability to outperform the single-node system. In contrast, the proposed SOE-MP technique, achieving 10 ps synchronization accuracy, enables distributed ISAC to fully exploit cooperative sensing gains, even under decentralized processing.

\section{Open Challenges and Future Directions} \label{sec::Challenge}
In this section, we highlight open challenges and future directions in network-level ISAC design and the practical implementation of distributed ISAC systems.
\subsubsection{Phase Synchronization for Coherent ISAC Cooperation}
Recent frameworks developed for distributed network-level ISAC cooperation, including cell-free ISAC systems, largely overlook the impact of phase misalignment. Achieving phase coherence is critical for maximizing cooperative gains in both communication and sensing \cite{yang2011phase, mudumbai2007feasibility}. However, realizing practical phase synchronization remains highly challenging. One primary difficulty stems from the requirement that perfect time and frequency synchronization must be established before phase alignment can be maintained. Furthermore, the use of independent oscillators at each node introduces phase noise, leading to time-varying phase offsets that hinder stable, long-term coherent operation. Despite these challenges, coherent transmission and processing offer significant performance advantages over noncoherent approaches, motivating continued efforts toward achieving phase-synchronized network-level ISAC designs.
\subsubsection{Low-Complexity Signaling Design and Processing}
Another major challenge in network-level ISAC design lies in the high complexity of centralized signaling and the substantial backhaul capacity requirements associated with centralized signal processing. While centralized coordination can theoretically achieve optimal performance for distributed ISAC systems, it often leads to degraded system efficiency and increased latency due to the extensive data exchange between nodes. Consequently, the practical deployment of networked ISAC demands decentralized (or local) signaling design at the transmitters and sensing signal processing at the receivers \cite{xiong2022distributed}. In this context, the integration of artificial intelligence (AI) techniques, such as model-based learning and federated learning, offers promising avenues to enhance decentralized signaling and processing. Additionally, semantic ISAC enables to achieve meaningful information transmission in a more efficient way as well as reduce burdens of the radar sensing data fusion \cite{10417099}. These approaches enable distributed nodes to achieve reliable cooperative gains while alleviating communication and computational overhead, thus improving the scalability and practicality of distributed ISAC systems.
\subsubsection{Secure Network-Level ISAC Designs}
ISAC systems face inherent challenges related to physical layer security (PLS) \cite{su2020secure}, while simultaneously offering new opportunities for developing PLS solutions \cite{su2023sensing, cao2024sensing}. Although initial research has explored secure link-level ISAC by jointly designing sensing and secure communication functions, the development of secure cooperative ISAC systems remains an open challenge. In particular, secure ISAC performance at the network level, where distributed nodes cooperate, has not yet been thoroughly investigated. Moreover, unlike traditional communication-only systems, ISAC introduces unique vulnerabilities associated with sensing security \cite{zou2024securing, ren2024secure}, which are still underexplored within current secure ISAC frameworks. Consequently, leveraging cooperation among distributed ISAC nodes may provide breakthroughs not only in enhancing data confidentiality but also in improving sensing security, opening new directions for secure network-level ISAC designs.

\section{Conclusion}
This paper has provided a comprehensive overview of network-level ISAC system design, an emerging research direction in the ISAC field. We reviewed recent advancements in network-level ISAC, focusing on cooperative schemes and distributed system architectures. The performance characteristics of networked ISAC systems were then analyzed, offering new insights into interference management and cooperative ISAC gain. Subsequently, we discussed distributed ISAC signaling designs across different cooperation levels, identifying key sensing and communication metrics critical for effective signaling optimization. In addition, synchronization challenges, a fundamental enabler for distributed ISAC systems, were addressed by reviewing state-of-the-art synchronization techniques tailored for distributed architectures. Finally, the paper outlined major open challenges and potential future research directions for distributed network-level ISAC design. As ISAC continues to evolve toward providing coordinated sensing and communication at unprecedented scales, we hope that the discussions and findings presented here will guide and inspire future advancements in network-level ISAC systems.

\ifCLASSOPTIONcaptionsoff
  \newpage
\fi



%
\bibliographystyle{IEEEtran}
\bibliography{IEEEabrv,references}

%








\end{document}